\documentclass[12pt]{article}
\usepackage{geometry}                
\usepackage[dvips]{graphicx}
\usepackage{amssymb}
\usepackage{amsmath}
\usepackage{epstopdf}
\usepackage{verbatim}

\def\ba{\begin{array}}
\def\ea{\end{array}}

\def\dalemb#1#2{{\vbox{\hrule height .#2pt
       \hbox{\vrule width.#2pt height#1pt \kern#1pt
               \vrule width.#2pt}
       \hrule height.#2pt}}}

\newcommand{\Tr}{{\rm Tr} }

\def\R{{{\Bbb R}}}

\def\Z{{{\Bbb Z}}}

\def\Im{{{\frak{Im}}}}
\def\Re{{{\frak{Re}}}}

\def\Op{{\mathcal{O}}}

\def\rh{{\tilde{r}}}
\def\th{{\tilde{t}}}
\def\phih{{\tilde{\phi}}}
\def\eh{{\tilde{\epsilon}}}
\def\omegah{{\tilde{\omega}}}

\newcommand{\be}{\begin{equation}}
\newcommand{\ee}{\end{equation}}
\newcommand{\bal}{\begin{align}}
 \newcommand{\eal}{\end{align}}
\newcommand{\ben}{\begin{equation*}}
\newcommand{\een}{\end{equation*}}
\newcommand{\bea}{\begin{eqnarray}}
\newcommand{\eea}{\end{eqnarray}}
\newcommand{\bean}{\begin{eqnarray*}}
\newcommand{\eean}{\end{eqnarray*}}
\newcommand{\bes}{\begin{subequations}}
\newcommand{\ees}{\end{subequations}}

\DeclareMathOperator{\sech}{sech}

\newcommand{\F}{{\mathcal F}}

\begin{document}

\begin{titlepage}
\bigskip
\rightline{}

\bigskip\bigskip\bigskip\bigskip
\centerline {\Large  {No Dynamics in the Extremal Kerr Throat }}

\bigskip\bigskip
\bigskip\bigskip

\centerline{\large  Aaron J. Amsel, Gary T. Horowitz, Donald Marolf, and Matthew M. Roberts}
\bigskip\bigskip
\centerline{\em Department of Physics, UCSB, Santa Barbara, CA 93106}
\bigskip
\centerline{\tt  amsel@physics.ucsb.edu, gary@physics.ucsb.edu}
\centerline{\tt marolf@physics.ucsb.edu, matt@physics.ucsb.edu}
\bigskip\bigskip

\begin{abstract}
Motivated by the Kerr/CFT conjecture, we explore solutions of vacuum general relativity whose asymptotic behavior agrees with that of the extremal Kerr throat, sometimes called the Near-Horizon Extreme Kerr (NHEK) geometry.    We argue that all such solutions are diffeomorphic to the NHEK geometry itself.  The logic proceeds in two steps. We first  argue that certain charges must vanish at all times for any solution with NHEK asymptotics.  We then analyze these charges in detail for linearized solutions.  Though one can choose the relevant charges to vanish at any initial time, these charges are not conserved.  As a result, requiring the charges to vanish at all times is a much stronger condition.  We argue that all solutions satisfying this condition are diffeomorphic to the NHEK metric.

\end{abstract}
\end{titlepage}

\tableofcontents

\setcounter{equation}{0}

\section{Introduction}

The near-horizon limit of an extreme Kerr black hole \cite{Zaslavskii:1997uu, Bardeen:1999px} is described by a geometry with isometry group $SL(2,\R) \times U(1)$ and is known either as the extremal Kerr throat \cite{Bardeen:1999px} or as the Near-Horizon Extreme Kerr (NHEK) geometry \cite{Guica:2008mu}.  As reviewed in section \ref{review} below, this spacetime has many properties in common with AdS${}_2 \times S^2.$ We will use the terms NHEK and extreme (or extremal) Kerr throat interchangeably throughout this work.

It was recently observed that, with appropriate boundary conditions, the asymptotic symmetry group of this spacetime contains a Virasoro algebra whose central charge is related via Cardy's formula to the entropy of the Kerr black hole \cite{Guica:2008mu}.  This observation led the authors of \cite{Guica:2008mu} to conjecture a full Kerr/CFT-correspondence, in analogy with the well-established AdS/CFT correspondence \cite{Maldacena:1997re}, in which the dynamics of a gravitational theory with such boundary conditions would be equivalent to some chiral 1+1 CFT.  As a result, there has been a great deal of recent interest in both the NHEK spacetime and its analogues in higher dimensions (see e.g. \cite{Azeyanagi:2009wf} and references therein).

However, many aspects of the proposed Kerr/CFT-correspondence remain deeply mysterious.  For example, in modular-invariant CFT's, Cardy's formula gives the density of states for large excitations above the ground state.  Thus, one expects the NHEK geometry to correspond to some highly excited state.  But then, what geometry is dual to the ground state of the CFT?  Because it is extremal, \cite{Guica:2008mu} suggested that the NHEK geometry itself be interpreted as a ground state, but further exploration of the dynamics may be enlightening.

A related question concerns the stability of the throat geometry.  Asymptotically flat Kerr black holes exhibit superradiance, meaning that certain modes of bosonic fields are amplified when they scatter off the black hole \cite{SuperRad}.  If one places such black holes inside a reflecting box \cite{PTBomb} (or in asymptotically anti-de Sitter space \cite{Cardoso:2004nk}), these modes continually reflect back and forth off of the black hole and the box wall.  Every cycle
amplifies the waves, leading to an exponential instability.  For non-extreme black holes, placing the box wall close enough to the horizon turns off this instability due to the fact that the very short wavelength modes are stable (as is typical for systems containing tachyons).   However, the infinite throat of the extreme Kerr black hole means that some instabilities can remain no matter how small a box is chosen\footnote{\label{f1}A more complete argument notes that the horizon-generating Killing field of a non-extreme Kerr black hole is timelike near the horizon, so that timelike observers sufficiently close to a non-extreme Kerr black hole can co-rotate with the black hole, out to the so-called velocity of light surface where such co-rotating observers must become null.  For positive-energy matter, this timelike Killing field defines a positive conserved quantity for excitations in the near-horizon region, ruling out instabilities.  In contrast, the horizon-generating Killing field of extreme Kerr is spacelike at all points near the equator outside the horizon, no matter how far one goes down the throat. As a result, the Frolov-Thorne vacuum \cite{Frolov:1989jh} for linear fields discussed in \cite{Guica:2008mu} is not well-defined in the extreme Kerr throat \cite{Kay:1988mu,Ottewill:2000qh,Ottewill:2000yr}.}. One therefore expects any attempt to separate the throat geometry from the asymptotically flat region to have instabilities.

These issues motivate a general study of solutions which agree asymptotically with the extremal Kerr throat.  The first steps are taken below.  We  analyze perturbations of the throat and their back-reaction, we classify general stationary axisymmetric asymptotically-NHEK solutions, and we study the near-horizon limits of perturbed non-extreme Kerr black holes.

Our final conclusions will turn out to be dominated by back-reaction effects.   To understand the importance of back-reaction, recall that non-linearities in gravity lead to two conceptually distinct effects.  The first is an effect on the dynamical evolution, while the second is an effect on the initial data that arises from the gravitational constraints. For example, in 3+1 dimensional asymptotically flat space the presence of any energy requires the initial data to contain a $1/r$ Coulomb tail.  As a result, initial data of compact support is generally not allowed, and it is the values of certain charges that determine the asymptotic fall-off properties of the gravitational field.  It is therefore important to determine whether the fall-off properties dictated by a given charge are compatible with the specified boundary conditions.  If not, then that charge must vanish for all solutions with the desired asymptotics.

Since this effect is fundamentally non-linear, the linearized equations of motion will generally admit solutions with non-zero charges, even after the desired fall-off conditions are imposed.   However, the back-reaction effects at the next order will satisfy the asymptotic conditions only for linearized solutions in which the relevant charges vanish.  For this reason, the conditions that these charges vanish are known as ``linearization-stability constraints."  The classic example of such constraints occurs for gravity on spacetimes with compact Cauchy slices \cite{BD,FM1,VMLS1,VMLS2,Arms,FM,AMM} (e.g., for periodic boundary conditions), though the same basic effect has recently been discussed in the context of chiral gravity \cite{Maloney:2009ck}.   In the asymptotically NHEK context, we will argue that linearization-stability constraints require all charges associated with the $SL(2,\R) \times U(1)$ isometries to vanish\footnote{See section \ref{LinStab} for subtle points involving boundary gravitons.}.

For this reason we begin with a discussion of back-reaction in section \ref{back}.  As a part of this analysis, we study near-horizon limits of asymptotically flat non-extreme Kerr solutions.  We seek scaling limits of such solutions which approach a given extreme Kerr throat at large distance and show that the charges of such scaling limits always vanish when measured relative to the relevant throat metric.    We also prove that the only stationary, axisymmetric, asymptotically-NHEK solution with a smooth horizon is the NHEK metric itself.  We interpret these results as evidence for the anticipated linearization stability constraints.

It then remains to impose these constraints on solutions to the linearized Einstein equations.  It is straightforward to analyze such solutions following the approach used by Teukolsky \cite{Teukolsky:1973ha,Press:1973zz,Teukolsky:1974yv} for asymptotically flat Kerr.  However, the analysis is rather cumbersome and is based heavily on both the Newman-Penrose formalism \cite{Newman:1961qr} (reviewed in appendix \ref{NP}) and the gravitational symplectic structure, technology which may be unfamiliar to many readers.  On the other hand, a massless scalar field provides a simple toy model of linearized gravity.  We therefore treat this model in great detail in section \ref{scalars}, before addressing linearized gravity itself in section \ref{grav}.  It will turn out that boundary conditions which conserve Klein-Gordon flux necessarily break some of the $SL(2,\R)$ symmetries.  As a result, some of the $SL(2,\R)$  charges are not conserved, and their vanishing imposes a separate condition on each Cauchy surface.  This breakdown of the initial value problem is an interesting departure from previous examples of linearization-stability constraints, and results in much stronger restrictions on the allowed solutions.   Within the class of generalized-Dirichlet boundary conditions, only the trivial solution $\Phi =0$ is compatible with this full set of constraints.

Section \ref{grav} is then dedicated to showing that solutions of the linearized Einstein equations behave in much the same way.  Although there are many interesting technical points in this analysis, the physics turns out to be identical to that of the much simpler scalar field.  The reader wishing to avoid the required formalism will miss little of the essential physics by skipping over section \ref{grav} on a first reading.  

In the bulk of this paper we use boundary conditions which require the metric to asymptotically approach that of the extreme Kerr throat.  Readers particularly interested in Kerr/CFT issues should note that the boundary conditions of \cite{Guica:2008mu} (which we call GHSS fall-off) are somewhat different.  We discuss the implications of our arguments for GHSS fall-off in section \ref{disc} and find that the results largely carry through.

\setcounter{equation}{0}

\section{Linearization-stability constraints for the extremal Kerr throat}
\label{back}

This section argues that the charges of any solution asymptotic to a given extremal Kerr throat are highly constrained.  This raises a number of linearization-stability issues which we will investigate further in sections \ref{scalars} and \ref{grav}.

After a brief review of the extremal Kerr throat in section \ref{review}, we begin to probe asymptotically NHEK solutions by analyzing general near-horizon limits of asymptotically-flat Kerr black holes (section  \ref{limits}).  If there exist non-extreme black hole solutions with extreme Kerr throat asymptotics, one might expect that such solutions could be constructed via such limits.  This would be in parallel with, for example, the construction of planar black holes in AdS${}_5$ from a similar scaling limit of black 3-branes.   Yet we find that charges of the limiting solutions always vanish relative to the appropriate NHEK background.  Furthermore, as shown in section \ref{StatAxi}, these turn out to be the only stationary axisymmetric asymptotically-NHEK solutions with smooth horizons.  Section \ref{LinStab} then interprets these results as evidence for linearization-stability constraints, commenting on certain subtleties involving boundary gravitons.

\subsection{Brief review of the extreme Kerr throat}
\label{review}

To orient the reader and establish conventions,
we begin by recalling how the extremal Kerr throat can be obtained as a scaling limit of the Kerr geometry \cite{Bardeen:1999px}.  The general Kerr metric is labeled by two parameters, a mass $M$ and angular momentum $J=Ma.$  The resulting black hole has temperature $\tilde T=\frac{\sqrt{M^2-a^2}}{4\pi M(M+\sqrt{M^2-a^2})}$ and entropy $S=2\pi M(M+\sqrt{M^2-a^2})$.  In Boyer-Lindquist coordinates $(\th,\rh,\theta,\phih)$, the metric takes the form
\be
ds^2=-e^{2\nu}d\th^2+e^{2\psi}(d\phih + A d\th)^2+\Sigma(d\rh^2/\Delta+d\theta^2) \,,
\ee
where
\be
\Sigma=\rh^2+a^2\cos^2\theta, \ \ \ ~\Delta=\rh^2-2M\rh+a^2 \,,
\ee
\be
e^{2\nu}=\frac{\Delta\Sigma}{(\rh^2+a^2)^2-\Delta a^2\sin^2\theta},~
\ \ \ e^{2\psi}=\Delta\sin^2\theta e^{-2\nu},~ \ \ \ 
A= -\frac{2M\rh a}{\Delta\Sigma}e^{2\nu}.
\ee
For the extremal solution $a=M,~S=2\pi M^2=2\pi J$.

Defining a one-parameter family of new coordinate systems
\be
\rh=M+\lambda r,~ \ \ \ \th=t/\lambda,~ \ \ \ \phih=\phi+t/2M\lambda \, , \label{nhekscaling}
\ee
and taking the scaling limit $\lambda\rightarrow 0$ yields
\be
ds^2=\left(\frac{1+\cos^2\theta}{2} \right)\left[ -f dt^2+dr^2/f+r_0^2d\theta^2\right]+\frac{2r_0^2\sin^2\theta}{1+\cos^2\theta}(d\phi+ r/r_0^2 dt)^2\, ,\label{nhekmetric}
\ee
with $r_0^2=2M^2$ and $f=r^2/r_0^2$.  This spacetime is known either as the extremal Kerr throat or as the Near-Horizon Extreme Kerr (NHEK) geometry.    For fixed $\theta$, the term in square brackets becomes the metric on $AdS_2$ in Poincar\'e coordinates.  As a result, we refer to  $(t, r, \theta, \phi)$ as Poincar\'e coordinates for the extremal Kerr throat.

The throat geometry inherits many properties from the above-mentioned AdS${}_2$.  For example, a geodesically complete spacetime can be obtained by
performing the coordinate transformation
\be
\label{PoincareToGlobal}
r= (1+y^2)^{1/2}\cos\tau+y,~t= \frac{(1+y^2)^{1/2}\sin\tau}{r},~\phi=\varphi+\log\left|\frac{\cos\tau+y\sin\tau}{1+(1+y^2)^{1/2}\sin\tau} \right| ,
\ee
which takes Poincar\'e AdS${}_2$ to the standard global coordinates on AdS${}_2$.  The result is again of the form (\ref{nhekmetric}) with $r$ replaced by $y$, $t$ replaced by $\tau$, $\phi$ replaced by $\varphi$, and $f = 1 + y^2/r_0^2$.  The analytic extension of the solution to the coordinate range $y, \tau \in (-\infty, \infty)$ is then geodesically complete.  This form of the metric is known as the NHEK geometry in global coordinates.  One notes that it has two boundaries, at $y = \pm \infty$.

The throat geometry also inherits the isometries of AdS${}_2$.  These are well-known to form an $SL(2,\R)$ algebra and are given by
\begin{equation}
\eta_{-1} = \left( \frac{1}{2 r^2} + \frac{t^2}{2} \right) \partial_t - tr \partial_r - \frac{1}{r}  \partial_\phi, \ \ \
\eta_0 =  t \partial_t - r \partial_r,  \ \
\eta_{1} =  \partial_t\, ,
\end{equation}
in Poincar\'e coordinates.  The Lie brackets of these vector fields satisfy
\begin{equation}
[\eta_0, \eta_{\pm 1}] = \mp \eta_{\pm 1}, \ \ \ [\eta_{1}, \eta_{-1}] = \eta_0.
\end{equation}
There is also one additional Killing field, $\xi_0 = \partial_\phi$, which commutes  with all $\eta_i$. For future reference we note that, using the particular diffeomorphism (\ref{PoincareToGlobal}), the global time translation is $\partial_\tau = \frac{1}{2}\eta_1 + \eta_{-1}$.

The $SL(2,\R)$ Killing fields are all closely related.  Indeed, the conjugacy class of a non-zero element of the Lie algebra of $SL(2,\R)$ is determined by its norm with respect to the Cartan-Killing metric, for which the associated quadratic form is $2\eta_1 \eta_{-1} - \eta_0^2$ (up to normalization), and a sign (future/past-directed) for null and timelike elements.  Thus, all null elements of the Lie algebra are related by conjugation and multiplication by a real number.  Furthermore, any Lie algebra element can be expressed as a linear combination of null elements, in the same way that one may choose a basis of null vectors for 2+1 Minkowski space.  Since the Poincar\'e time translation $\eta_1$ is a null element, it follows that one may think of the general $SL(2,\R)$ vector field as a linear combination of isometries, each of which is just the Poincar\'e time translation.  Because the isometries are so closely related, we shall take care to label the various charges by the relevant Killing fields; e.g., we shall speak of $Q_{\eta_i}$ and $Q_{\xi_0}$.

Despite the many similarities of (\ref{nhekmetric})  to AdS${}_2\times S^2$, there are also some important differences.  For example,  the Poincar\'e time translation becomes spacelike near the equator   ($\theta = \pi/2$) of the sphere\footnote{This is just the statement mentioned in footnote \ref{f1} that the horizon-generating Killing field of extreme Kerr is spacelike near the equator, even in the near-horizon region.}.  The time translation $\partial_\tau$ associated with global coordinates also becomes spacelike near the equator for large $r$.  In fact, any linear combination of the above Killing fields becomes spacelike in some region of the spacetime.

\subsection{Throat limits of general Kerr black holes}
\label{limits}

As noted in the introduction, we wish to argue that solutions approaching (\ref{nhekmetric}) at large $r$ are highly constrained.    We begin by seeking additional solutions which can be obtained as scaling limits of asymptotically  flat non-extreme black holes.  In parallel with the AdS${}_2\times S^2$ case studied in \cite{Maldacena:1998uz}, every such metric turns out to be diffeomorphic to the original extreme throat (\ref{nhekmetric}).   These scaling limits will also lead to a physical argument for linearization-stability constraints in section \ref{LinStab}.

The asymptotically flat Kerr black holes with which we begin have two non-zero charges, associated with time-translations and rotations.  We will use conventions in which $Q_{\partial_{\tilde t}} = -M$ and $Q_{\partial_{\tilde \phi}} = J = aM$, so that  $Q_\xi$ is a linear function of $\xi$. We consider one-parameter families  of black holes specified by giving $M,J$ as functions of the parameter ($\lambda$).  We take the scaling limit to be given by (\ref{nhekscaling}) up to subleading corrections.    In order to develop a throat region, we require that $J$ approach some extremal values  $J \rightarrow M^2 \rightarrow r_0^2 /2$  for some $r_0 > 0$ as $\lambda \rightarrow 0$.    It is also useful to introduce a non-extremality parameter $\eh$ defined by
\be
\eh^2=M^2-J^2/M^2,
\ee
so that $\eh \rightarrow 0$.

We require the coordinate transformation to agree with (\ref{nhekscaling}) at leading order.    In particular, we have
\be
\label{tscale}
\th \rightarrow t/\lambda
\ee
where the arrow ($\rightarrow$) indicates that we allow arbitrary subleading corrections.  The rate at which $\tilde \epsilon$ must vanish can then be determined by noting that any solution with a smooth horizon must have finite temperature $T$ with respect to the rescaled time coordinate $t$.
Since horizon temperature can be related to the period of imaginary time, (\ref{tscale}) is enough to determine the scaling behavior of the temperature:
\be
\label{T1}
T \rightarrow  \frac{\tilde T}{\lambda} = \frac{\tilde \epsilon/\lambda}{4 \pi M(M+\tilde \epsilon)} \rightarrow \frac{\tilde \epsilon/ \lambda }{2 \pi r_0^2}.
\ee
Thus, in order to obtain a finite temperature,
we must require $\tilde \epsilon \rightarrow  \lambda \epsilon$ for some fixed $\epsilon$.
Using $J^2 = M^4 - \tilde \epsilon^2 M^2$, it follows that
\be
\label{Jas}
J(\lambda) \rightarrow M^2(\lambda) - \frac{\lambda^2 \epsilon^2}{2};
\ee
i.e., that the effects of non-extremality enter only at second order in $\lambda$.

We can now use (\ref{Jas}) to show that charges associated with the limiting solution do not depend on the non-extremality parameter $\epsilon$.  The point is that, since solutions are invariant under both $t$- and $\phi$-translations, the charges can be expressed in terms of Komar integrals and so may be evaluated on any closed two-surface in the geometry; i.e., they may be evaluated at any finite position without taking the limit $r \rightarrow \infty$.   As a result, the charges of the asymptotically NHEK solutions obtained from our scaling limits must be given by limits of the asymptotically flat charges.   Now, it is important to recall that the charges depend on the choice of vector field.  Nevertheless, we have

\be
\label{c1}
{\partial_\phi} \rightarrow  {\partial_{\tilde \phi}} ,  \ \ \
{\partial_t} \rightarrow \frac{1}{\lambda} \left( \partial_{\tilde t}   + \frac{1}{2M}  {\partial_{\tilde \phi}}  \right) .
\ee
Since these expressions contain only one factor of $\lambda^{-1}$ while $\epsilon^2$ appears in (\ref{Jas}) multiplied by $\lambda^2$, the $\lambda \rightarrow 0$ limit of any charge is independent of $\epsilon.$  Now, applying the scaling limit directly to the charge $Q_{ {\partial_t}}$ yields an infinite result as $\lambda \rightarrow 0$ which must be regulated by subtracting the charge of some reference scaling limit.   But even such regulated results cannot depend on $\epsilon.$  It follows that scaling limits of non-extreme Kerr black holes yield precisely the same charges $Q_{{\partial_\phi}}$ and $Q_{ {\partial_t}}$ as do scaling limits of extreme Kerr black holes, independent of the temperature $T$. 

When $\epsilon \neq 0$, the actual metric obtained from (\ref{nhekscaling}) in the $\lambda \rightarrow 0$ limit is not (\ref{nhekmetric}), but
instead
\begin{eqnarray}
\label{finiteT}
ds^2 &=&
\frac{1+\cos^2\theta}{2}\left(- \frac{r_T(r_T-2 k_T)}{r_0^2}dt_T^2+\frac{r_0^2dr_T^2}{r_T(r_T-2 k_T)}+r_0^2d\theta^2 \right) \cr &+&\frac{2 r_0^2\sin^2\theta }{1+\cos^2\theta}\left(d\phi_T + \frac{r_T-k_T}{r_0^2}dt_T\right)^2,
\end{eqnarray}
where $r_0^2=2M^2$ as before, and $k_T= \pi r_0^2 T.$ As in (\ref{T1}),  $T$ is the Hawking temperature of the resulting black hole. One might also attempt to obtain more general metrics by modifying sub-leading terms of (\ref{nhekscaling}), but we have found no other useful limits of this form. Furthermore, any such limits would be restricted by the uniqueness results of section \ref{StatAxi}.

\begin{figure}
\begin{center}
\includegraphics[width=8cm]{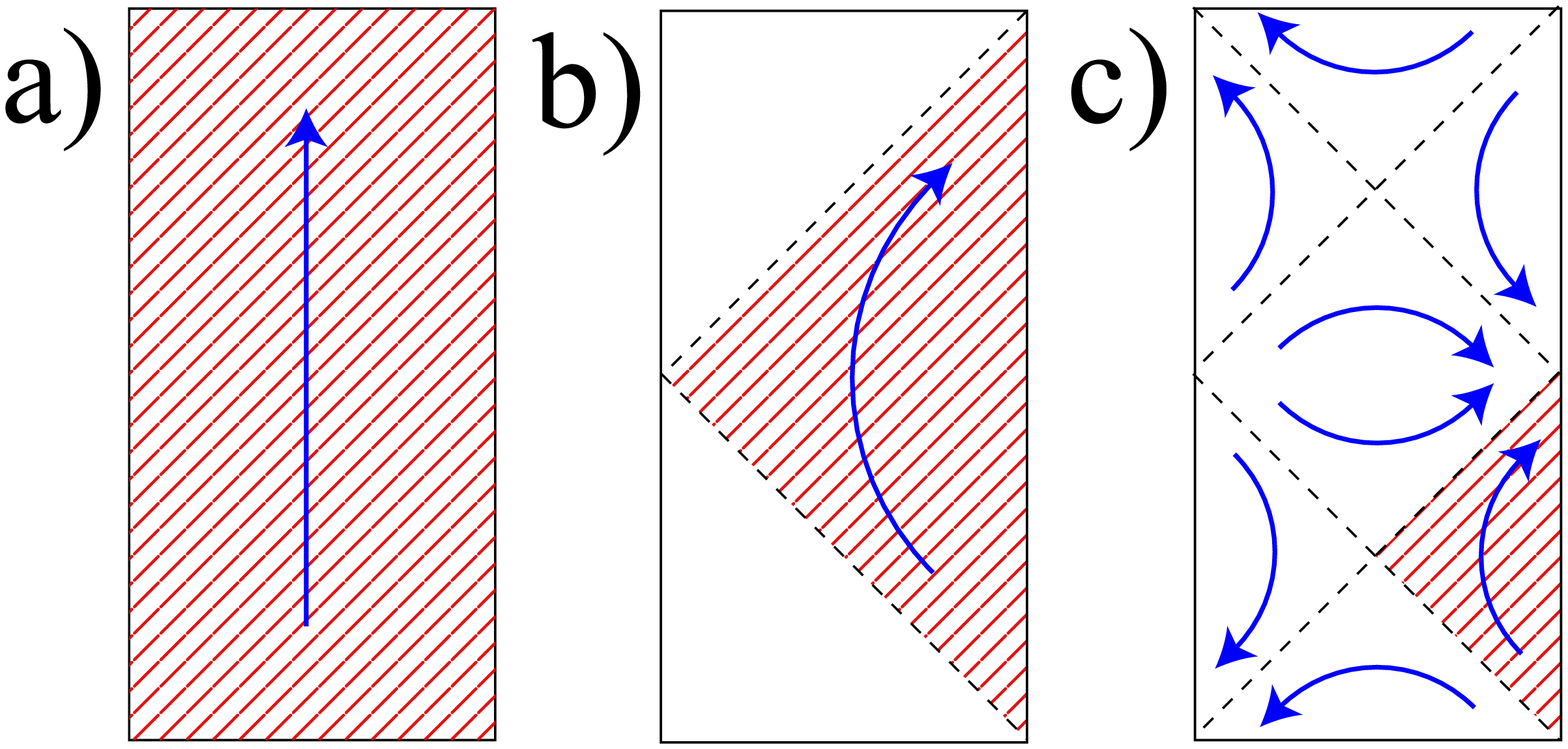}
\end{center}
\begin{caption}
{Penrose diagrams of AdS${}_2$ showing coordinate patches covered by (a) global coordinates, (b) Poincar\'e coordinates, (c) finite temperature coordinates.  In each case the associated time translation and any horizons are shown.  \label{AdS2}}
\end{caption}
\end{figure}

In (\ref{finiteT}) we have renamed the coordinates $(t_T, r_T, \phi_T)$ due to the fact that, as in the $AdS_2\times S^2$ case \cite{Maldacena:1998uz}, the diffeomorphism
\be
\label{TtoPoincare}
r_T=T(r_0^2-rt ), \ \ ~e^{t_T/T}=\frac{r_0 r}{\sqrt{r^2 t^2-r_0^4}}, \ \ ~\phi_T =\phi + \log\sqrt{1-2k/r_T}
\ee
takes (\ref{finiteT}) to (a subset of) the standard Poincar\'e NHEK solution (\ref{nhekmetric})\footnote{We thank Geoffrey Comp\`ere for helping to discover this.}.  As a result, one may view (\ref{finiteT}) as the NHEK geometry written in terms of a one-parameter family of coordinate systems $(t_T, r_T, \phi_T, \theta)$.        In order to help the reader visualize the various systems of coordinates,  Figure \ref{AdS2} displays the global, Poincar\'e, and finite-temperature time-translations and the associated horizons on a Penrose diagram of AdS${}_2$.

For later purposes we note  that $\frac{\partial}{\partial t_T}$ maps to $-T^{-1} \eta_0$ under (\ref{TtoPoincare}); i.e., to a spacelike element of the $SL(2,\R)$ isometry algebra.    Since any other spacelike element is related by conjugation, and since spacelike elements form a basis for the Lie algebra of $SL(2,\R)$, it follows that one may also think of a general element of $SL(2,\R)$ as simply a linear combination of finite-temperature time-translations.

Although we derived the condition $\tilde \epsilon \rightarrow  \lambda \epsilon$ by requiring the horizon to be smooth, it is straightforward to check that taking $\tilde \epsilon$ to vanish more slowly leads to a metric which diverges everywhere, and not just at the horizon.  Thus, even without requiring regularity in the interior, we find that all scaling limits of Kerr which agree asymptotically with (\ref{nhekmetric}) have charges
$Q_{\partial_t}$ and $Q_{\partial_\phi}$ identical to (\ref{nhekmetric}).  Applying the technology of e.g. \cite{Barnich:2001jy} to (\ref{finiteT}), one readily checks that this statement holds for the other $SL(2,\R)$ charges as well.

\subsection{Stationary axisymmetric solutions}
\label{StatAxi}

As advertised above, we  now classify {\it all} stationary,  axisymmetric, asymptotically-NHEK vacuum solutions having smooth horizons. Since it is clear that in (\ref{nhekmetric}), the norm of $\partial/\partial t$ changes sign even asymptotically as one changes the polar coordinate $\theta$, we will define ''stationary, axisymmetric'' in the context of asymptotically-NHEK geometries to simply mean geometries with $R\times U(1)$ isometry group. They are given by the following uniqueness theorem:

\bigskip
{\bf Theorem:} Any asymptotically NHEK vacuum solution with an $R\times U(1)$ isometry group and a smooth horizon (either extremal or non-extremal) is diffeomorphic to the NHEK solution itself.
\bigskip

{\it Proof:} For non-extreme horizons, our proof will be similar to proofs of the uniqueness of the (asymptotically flat) Kerr black hole. The only change is in the boundary conditions at infinity. (We will see that the proof is actually somewhat easier with NHEK boundary conditions.) We follow the approach in \cite{Hollands:2007aj,Hollands:2008fm} which is based on earlier work by Mazur \cite{Mazur:1982db}. Since the argument is identical to these earlier proofs except for the asymptotic boundary conditions, we will just give the main ideas. For technical details, we refer the reader to \cite{Hollands:2007aj,Hollands:2008fm,Chrusciel:2008js}.  The proof for extremal horizons that is given here can be extended to prove the uniqueness of asymptotically flat extremal Kerr black holes. Details will be given elsewhere \cite{us}.

 Stationary, axisymmetric metrics can be written in the Papapetrou form,
\be\label{papapetrou}
 ds^2 = -{\rho^2 \over F} dt^2 + F(d\phi + A dt)^2 + e^{2\nu}(d\rho^2 + dz^2) \,,
 \ee
 where $F, A,  \nu$ are functions of $\rho$ and $z$ only. Regularity along the axis requires that $F$ vanish as $\rho^2$ and no faster. Given a solution for $F$ and $A$, $\nu$ is then determined in terms of them by first order equations. Rather than work with $A$, it is convenient to work with the potential $\chi$ for the twist of the $\xi_0 = \partial_\phi$ Killing field:
 \be
 d\chi = *(\xi_0 \wedge d\xi_0).
 \ee
 A key role in the proof will be played by the following  $2 \times 2$ matrix constructed from the norm and twist of $\xi_0$:
 \be
 \Phi = {1\over F}\left(\begin{array}{cc} 1 &  - \chi \\-\chi & F^2 + \chi^2\end{array}\right) .
 \ee
 For the NHEK geometry, the twist potential is
\be
\chi_{NHEK} = -{4\cos\theta\over 1+\cos^2\theta} r_0^2
\ee
and $\Phi$ is a function of $\theta$ only:
\be\label{phinhek}
\Phi_{NHEK} = {1\over 2r_0^2\sin^2\theta} \left(\begin{array}{cc}{1+\cos^2\theta} & {4r_0^2\cos\theta} \\{4r_0^2\cos\theta}  & {4r_0^4(1+\cos^2\theta)}\end{array}\right).
\ee
This is true for both the Poincar\'e (\ref{nhekmetric}) and finite temperature (\ref{finiteT}) forms of the NHEK solution. The relation between $\theta$ and $(\rho,z)$  depends on the temperature, as we discuss below.

For a general solution, the matrix $\Phi$ is symmetric, has positive trace and unit determinant. It is therefore positive definite and can be written as $\Phi = S^T S$ for some matrix $S$  with $\det S = 1$. The equation satisfied by $\Phi$ is most easily expressed by viewing $\rho$ and $z$ as cylindrical coordinates in an auxiliary flat Euclidean $\R^3$, with derivative $\nabla_i$. Viewing $\Phi$ as a rotationally invariant matrix in this space, the vacuum Einstein equation implies
 \be\label{Phieq}
\nabla^i (\Phi^{-1} \nabla_i \Phi) = 0 \,,
\ee
where this equation holds everywhere except possibly the axis $\rho =0$.

Suppose that we have two axisymmetric solutions $\Phi_1$ and $\Phi_2$ to this equation.  Define
\be
\sigma = \Tr(\Phi_2 \Phi_1^{-1}) - 2,
\ee
or, in terms of the norm and twist of $\xi_0$,
\be
\label{sigmaaxis}
\sigma = {(\chi_1- \chi_2)^2 + (F_1 - F_2)^2\over F_1 F_2} \, .
\ee
Thus $\sigma \ge 0$. If in addition we set
\be N_i = S_2(\Phi_2^{-1} \nabla_i\Phi_2 - \Phi_1^{-1}\nabla_i\Phi_1) S_1^{-1},
\ee
then $\sigma$ satisfies the following ``Mazur identity"
\be \label{mazur}
\nabla^2 \sigma = \Tr(N_i^T N^i),
\ee
where again this equation holds everywhere except possibly the axis.   Note that the right hand side is nonnegative.

The requirements that $\nabla^2 \sigma \ge 0$ and $\sigma \ge 0$ impose strong constraints on $\sigma$. If we can show that $\sigma $ is globally bounded on $\R^3$ (including the axis) and vanishes at infinity then it follows  that $\sigma = 0$ everywhere \cite{Weinstein}. This, in turn, implies that
$\Phi_1 = \Phi_2$ and hence the two solutions agree.

To show that $\sigma$ is indeed bounded and vanishing at infinity, we  consider  the cases of degenerate and nondegenerate horizons separately.
\bigskip

{\it Case 1: Nondegenerate horizon}

The finite temperature form of the NHEK geometry (\ref{finiteT}) depends on two parameters,  $(r_0, k_T)$ related to the angular momentum $J$ and Hawking temperature $T$ by $r_0^2 = 2|J|$ and $k_T = \pi r_0^2 T$. It can be put into the Papapetrou form by defining
\be
\rho = \sqrt{r_T(r_T - 2k_T )} \sin\theta,\quad z = (r_T-k_T) \cos\theta \,.
\ee
The horizon at $r_T = 2k_T$ gets mapped into the ``rod" $\rho =0$ and $-k_T \le z \le k_T$ corresponding to the closure of the region along the axis for which $F = |\xi_0|^2 > 0$. Now consider a general nondegenerate (stationary, axisymmetric) horizon. Since the two-plane spanned by $\partial_t$ and $\partial_\phi$ becomes null at the horizon, $\rho$ must vanish there.  The horizon thus defines a similar rod, by the closure of the finite subset of the axis where $F > 0$. Now let $\Phi_1$ be any axisymmetric solution to the $\Phi$ equation (\ref{Phieq}) with a nondegenerate horizon and set $\Phi_2$ to be the NHEK solution with the same length rod and same angular momentum. On the horizon, $\Phi$ is finite away from the endpoints of the rod, so $\sigma $ is bounded. At infinity, since both solutions approach $\Phi_{NHEK}$ which itself is bounded, $\sigma \rightarrow 0 $ \footnote{In the usual asymptotically flat case, $\Phi$ is not bounded at infinity and this step requires more work. The boundary condition we need here is simply that $F = F_ {NHEK} +$ subleading terms, and $\chi = \chi_{NHEK}  +$ subleading terms, so that  $\Phi = \Phi_{NHEK} + $ terms that vanish asymptotically. The astute reader may note that this is a slightly stronger boundary condition than that used in \cite{Guica:2008mu}, a point which will be discussed further in section \ref{disc}. }. Finally, it remains to check the behavior of $\sigma$ on the rotation axis. This is a potential problem since $\Phi$ itself diverges there like $1/\rho^2$. However one can show that $\sigma$ remains bounded as follows: Since the rotational Killing vector $\xi_0 $ vanishes on the axis, its twist vector vanishes there and hence the twist potential $\chi$ is constant along the axis. Since the axis goes out to infinity, $\chi$ must have the value determined by the asymptotic NHEK geometry and hence $\chi_1 = \chi_2 $ on the axis.  Since $d\chi $ must vanish on the axis, $\chi_1 - \chi_2 =O(\rho^2)$ near the axis.  Since $F$ vanishes as $\rho^2$ and no faster near the axis, (\ref{sigmaaxis}) shows that $\sigma$ indeed remains bounded near the axis. Hence $\sigma$ is globally bounded on $\R^3$ and vanishes at infinity. Therefore it must vanish everywhere and $\Phi_1 = \Phi _2$. This shows that the only stationary axisymmetric asymptotically NHEK solution with a nondegenerate horizon is the NHEK geometry itself.
\bigskip

{\it Case 2: Degenerate horizon}

It was shown in \cite{Kunduri:2008rs} that the near horizon geometry of an extremal rotating vacuum black hole is given by the NHEK solution. To see the connection between this statement and the theorem we wish to prove, consider the NHEK solution in the form (\ref{nhekmetric}). The degenerate horizon is at $r=0$ and the asymptotic region is $r = \infty$. The result in \cite{Kunduri:2008rs} shows that near $r=0$ a general solution must agree with NHEK, and our boundary condition requires that at large $r$ the general solution must again agree with NHEK. In fact, in these two limits the $r_0$ parameters in the NHEK solutions must agree since they are determined by the angular momentum $J$, which can be computed at any $r$ via a Komar integral.   Nonetheless, a priori, there could be many solutions with different radial dependence which interpolate between these two regimes.

To put (\ref{nhekmetric}) into standard form, note that the $(r,\theta)$ part of the metric is conformal to $dr^2 + r^2 d\theta^2$, so if one sets
\be
\rho = r\sin\theta, \quad z=r\cos\theta,
\ee
then (\ref{nhekmetric}) takes the form (\ref{papapetrou}). In other words, in this case, the radial coordinate in (\ref{nhekmetric}) is the standard radial coordinate in the auxiliary space $\R^3$. In particular,  the horizon corresponds to the origin of this space.  Note that $\Phi_{NHEK}$ has a direction dependent limit there. However,  since $\Phi_1$ and $\Phi_2$ must agree with $\Phi_{NHEK}$ near $r=0$, $\sigma \rightarrow 0$ at $r=0$. Similarly, $\sigma \rightarrow 0$ at large $r$ by our boundary conditions. Along the axis, $\sigma$ remains bounded by the essentially same argument as in the nondegenerate case, though see \cite{us} for additional discussion of subtleties where the axis meets the horizon. Hence, $\sigma$ is globally bounded and must vanish. This completes the proof.

 It is likely that one can also prove uniqueness of the NHEK solution in global coordinates, where there are no horizons. However in this case, there are two asymptotic regions and the coordinate $\rho$ is no longer a good global coordinate. A proof would require further study of  solutions near the critical points of $\rho$.

\subsection{Linearization-stability constraints}

\label{LinStab}

Let us now pause to contemplate the physical implications of our results.  We begin by returning to scaling limits of asymptotically flat vacuum solutions, but this time we consider a nearly-extreme Kerr black hole perturbed by a small amount of gravitational radiation.  Here it is useful to suppose that we work in some coordinate patch that covers both sides of the Einstein-Rosen bridge and thus includes two asymptotic regions.  Recall that the charges can differ in the two asymptotic regions, with this difference being governed by the details of the perturbation.

Suppose that, at some time $\tilde t =0$, the gravitational radiation is confined to a region deep within the throat.  For simplicity we assume that the initial data agrees exactly with that of Kerr outside of a compact region\footnote{\label{K2} The results of \cite{Corvino:2003sp} state that, given essentially arbitrary asymptotically flat initial data (call it $\Sigma$) and a compact set $K_1$, one can construct another initial data set $(\hat \Sigma)$ which agrees precisely with $\Sigma$ inside $K_1$, but such that outside of some larger compact region $K_2$ the new data $\hat \Sigma$ is exactly that of Kerr for some $M, J$.  Moreover, one can choose $\hat \Sigma$ such that $M, J$ are arbitrarily close to the mass and angular momentum of $\Sigma$.   While \cite{Corvino:2003sp}  gives no bounds on the size of $K_2$, it is natural to suppose that $K_2$ can be taken to lie far down the throat in any limit in which $K_1$ is also pushed far down the throat.  }.   Under what conditions can one find a scaling limit of this solution asymptotic to (\ref{nhekmetric})?  For simplicity, let us assume that the masses and angular momenta of the two asymptotic regions agree (up to appropriate signs), as is the case when there is a symmetry that exchanges the asymptotic regions.  In this case, our scaling limit must approach (\ref{nhekmetric}) with the same value of $r_0$ as  either $y \rightarrow  \infty$ or $y \rightarrow -  \infty$.

It is clear that  one must push the radiation far down the throat in the desired limit.  Thus, the throat region outside the radiation becomes approximately that of some Kerr black hole.  So, outside the radiation, the problem reduces to the one studied above.   However, the diffeomorphism invariance of gravitational theories requires that the charges be given entirely by boundary terms.  As a result, the charges should depend only on the metric in this exterior region and, by the argument above, must agree with those of (\ref{nhekmetric}) for the given value of $r_0$.  If no radiation is present, this is one of the limits just discussed. The result is therefore a metric of the form (\ref{finiteT}) for some $T >0$ or (\ref{nhekmetric}) for $T=0$.

Although it is convenient to think of gravitational charges as boundary terms, they may also be expressed as bulk integrals by using the gravitational constraints.  Thus, the condition that the same charges be obtained at each boundary (and, furthermore, that they agree with those of (\ref{nhekmetric})) restricts the radiation allowed in the throat. We can summarize this restriction simply  in the limit where the perturbation is small, so that it defines a solution to the linearized Einstein equations.  The linearized theory admits a conserved charge $Q^{lin}_\xi$ for each isometry $\xi$ of the background.  Consider in particular the time translation $\partial_{t_T}$ and rotation $\partial_{\phi_T}$ of (\ref{finiteT}). With appropriate choices of signs and in the perturbative limit, each linearized charge is the difference between two asymptotic charges (one on each boundary).    It follows that the small-amplitude limit of such solutions can be obtained from a scaling limit of asymptotically flat solutions with vanishing linearized charges $Q^{lin}_{\partial_{t_T}}$ and $Q^{lin}_{\partial_{\phi_T}}$.

The above scaling limits suggest that asymptotically NHEK perturbations may be subject to the linearization-stability constraints $Q^{lin}_{\partial_{t_T}} = - T^{-1} Q^{lin}_{\partial_{\eta_0}} =0$ and $Q^{lin}_{\partial_{\phi_T}} = 0$.  Since every $SL(2,\R)$ element is a linear combination of finite-temperature time-translations,  the charges $Q^{lin}_{\eta_i}$ should vanish for the other $SL(2,\R)$ generators as well.     One may also expect such results based on the analogy with AdS${}_2 \times S^2$.  It is easy to find linearized solutions for, say, massive scalar fields on AdS${}_2$.  However,  the results of \cite{Maldacena:1998uz} show that these solutions cannot be extended to non-linear scalar-Einstein-Maxwell solutions asymptotic to AdS${}_2 \times S^2$ unless a certain integral of the stress tensor vanishes.  This result appears to be closely related to the Birkhoff-like theorem \cite{MTW} stating that the only spherically symmetric solutions of 4d Einstein-Maxwell theory are the Reissner-Nordstrom family of solutions and AdS${}_2\times S^2$.

One might ask if there can be more general  asymptotically-NHEK solutions which are simply not given by the scaling limits discussed above.  While a complete treatment of this loophole is beyond the scope of the current work, we showed in section \ref{StatAxi} that all stationary, axisymmetric solutions asymptotic to (\ref{nhekmetric}) and having a regular horizon are diffeomorphic to either (\ref{nhekmetric}) or (\ref{finiteT}) for some $T$.  This result supports the conjecture that any solution asymptotic to (\ref{nhekmetric}) is diffeomorphic to a solution whose charges are determined by $r_0$.    Such a conjecture would in turn imply linearization-stability constraints for the $SL(2,\R) \times U(1)$  charges.   In the rest of this work, we will assume linearization-stability constraints of this form.

The astute reader may note that our discussion thus far has ignored all issues related to boundary gravitons. We now pause briefly to address such concerns.    Recall that boundary gravitons are excitations generated by diffeomorphisms that, because of the particular boundary conditions imposed in a given problem, are not pure gauge; i.e., that they are non-degenerate directions of the symplectic structure.  However, their restriction to any subset of the spacetime which does not reach the boundary is pure gauge.  As a result, our scaling argument above forbade non-trivial boundary gravitons through the requirement that the solution be precisely (\ref{nhekmetric}) outside some compact region at $t=0$. Note also that section \ref{StatAxi} classified solutions only up to diffeomorphisms, and so placed no restrictions on boundary gravitons.

As a result, our linearization-stability constraints need not apply to boundary gravitons, which must be analyzed separately. The properties, and in fact the very existence, of boundary gravitons depend on the particular choice of boundary conditions.  Here we simply note that, as discussed in that reference
\footnote{
In addition to the vector fields stated in eqn (5.2) of \cite{Guica:2008mu}, the GHSS
fall-off conditions given by eqn (5.1) of \cite{Guica:2008mu} are also invariant under
diffeomorphisms associated with vector fields which asymptote to $y \partial_y-\tau\partial_\tau$.
After the additional extremality condition $Q_{\partial_\tau} = 0$,
the charge associated to $y \partial_y-\tau\partial_\tau$ vanishes identically, and so it is not an
element of the asymptotic symmetry group.
}
, the Kerr/CFT boundary conditions of \cite{Guica:2008mu} lead only to boundary gravitons associated with the Virasoro generators of \cite{Guica:2008mu}.  Since $\eta_1$ and $\eta_0$ commute with the Virasoro generators, it follows that  the boundary gravitons do not carry either of the charges $Q_{\eta_0}$ or $Q_{\eta_1}$.  One can check that they also carry no $Q_{\eta_{-1}}$ charge, and that their $U(1)$ charges are determined by their higher Virasoro charges.  Thus, one may also think of these boundary gravitons as satisfying a slightly modified $U(1)$ linearization-stability constraint.     It will not be necessary to distinguish below between the simple $U(1)$ linearization-stability constraint (which requires the linearized charge to vanish) and the modified constraint (which determines the $U(1)$ charge in terms of the Virasoro charges) satisfied by boundary gravitons.

\setcounter{equation}{0}

\section{Linear scalar fields}
\label{scalars}

Having motivated the existence of linearization-stability constraints, we now investigate the extent to which they are consistent with the dynamics of linear fields.    Our goal is to study linearized gravity.  However, it is useful to first consider linear scalar fields propagating on (\ref{nhekmetric}).  In a combined scalar plus gravity theory, the linearization-stability constraints could receive contributions from both fields.  But it is interesting to consider a toy model in which we impose $Q^{lin}_{\eta_i} =0$ and $Q^{lin}_{\xi_0} =0$ on the scalar field alone.  We will see in section \ref{grav} that this toy model captures all the essential physics of gravitational perturbations, though much less technology is required to analyze the scalar case.

\subsection{The linear scalar wave equation}

We begin with the massless Klein-Gordon equation $\nabla^2\Phi=0$.  (This $\Phi$ should not be confused with the matrix $\Phi$ of section \ref{StatAxi}). As noted in \cite{Bardeen:1999px}, this equation is separable in the extreme Kerr throat.  It is convenient to set $r_0 =1$.  For $\Phi=e^{-i\omega t}e^{+im\phi}\Theta(\theta)R(r)$, one then obtains a radial equation

\be (f R')'+\left(\frac{(\omega+m r)^2}{f}+m^2-K\right)R=0,\label{NHEKwave} \ee
where in Poincar\'e coordinates $f=r^2$ and in global coordinates $f=1+r^2$.  The corresponding angular equation is

\be 
\frac{1}{\sin\theta}(\sin\theta~\Theta')'+\left(K-\frac{m^2}{\sin^2\theta}-\frac{m^2}{4}\sin^2\theta\right)\Theta=0, \label{NHEKscalarangular} 
\ee
whose solutions are deformations of standard spherical harmonics.  In particular, for $m=0$ we have axisymmetric spherical harmonics $\Theta=Y_{\ell,m=0}$ so that $K_{\ell,m=0}=\ell(\ell+1).$   For other values of $m$, the spectrum of $K$ will again be discrete and we can label the eigenvalues $K_{\ell, m}$.  It is natural to take $\ell$ to be an integer satisfying $\ell \ge m$ as for standard spherical harmonics.  Adding a (mass)${}^2$ term (of either sign) to the Klein-Gordon equation would merely shift the value of $K$ which, as we will see, leaves the qualitative behavior unchanged.

For general $m \neq 0$ the spectrum of $K$ must be found numerically.  Some analysis for small $m$ was performed in \cite{Bardeen:1999px}.  However, it is useful to consider solutions with large angular momentum localized near the equator $\theta = \pi/2.$  In this regime one may expand (\ref{NHEKscalarangular}) around $\theta=\pi/2$ to find
\be
\Theta''+(K-\frac{5m^2}{4})\Theta+\Op(\theta-\pi/2)=0,
\ee
whose solutions are just plane waves $\Theta \approx e^{ip(\theta-\pi/2)}$ with
$K \approx p^2+\frac{5m^2}{4}$.  While exact eigenstates of (\ref{NHEKscalarangular}) may not be localized near the equator, it is clear that one can find wavepackets with
$\langle K\rangle \approx \frac{5m^2}{4}+ \Op(m)$ so that the spectrum of $K$ must contain eigenvalues of this form. Furthermore, for given $m \gg 1$ these should be the lowest eigenvalues $K$.  In other words, we conclude

\be
\label{l=m}
K_{\ell =m ,m}=\frac{5}{4}m^2+\Op(m).\ee

The radial equation proves to be easier to study analytically. We will discuss the exact solutions shortly, but it is useful to first note that in the large $r$ limit one finds power law solutions with \be R\approx r^\Delta,~\Delta=- \frac{1}{2}\pm(K-2m^2+\frac{1}{4})^{1/2}.\ee

It is here that one finds an interesting difference between (\ref{nhekmetric}) and AdS${}_2 \times S^2.$  Note that the quantity $K-2m^2$ plays the role of an effective mass on the $r,t$ plane.  On AdS${}_2 \times S^2$, Kaluza-Klein reduction of a massless scalar leads to a tower of states with positive (mass)${}^2$ in AdS${}_2$.  In the NHEK geometry, while $K-2m^2$ can be arbitrarily positive for, say, axisymmetric modes, we see from (\ref{l=m}) that it can also be arbitrarily negative for maximally rotating modes.  As a result, reduction on the sphere effectively leads to a bi-directional tower of states which includes arbitrarily tachyonic masses.  In particular, for $K<K_{crit}=2m^2-1/4$, the exponent $\Delta$ becomes complex so that $R(r)$ is oscillatory.  In anti-de Sitter space, this happens only for scalars with masses below the Breitenlohner-Freedman (BF) bound \cite{Breitenlohner:1982bm} and generally leads to instabilities. We will see that this is also true of oscillatory modes in the NHEK geometry.

To discuss exact solutions to (\ref{NHEKwave}), it is useful to write $R \approx r^{- \frac{1}{2} + \mu}$ where 
\be \mu^2 = K- 2m^2 + 1/4.
\ee
We work in the Poincar\'e patch for simplicity, and to aid comparison with the graviton case. A brief discussion of scalars in global coordinates is provided in appendix \ref{global}. The precise form of the solution and spectrum differ slightly in these two cases but the physics is essentially the same.

Rewriting (\ref{NHEKwave}) in terms of $\mu$ and $m$ yields
\be 
(r^2 R')'+\left(\frac{1}{4}-\mu^2+\frac{\omega^2}{r^2}+\frac{2 m \omega }{r}\right)R=0,
\ee
which,
under the variable change $z=-2\omega i/r$, becomes Whittaker's equation:
\be 
R''+\left(-\frac{1}{4}+\frac{i m}{z}+\frac{1/4-\mu^2}{z^2}\right)R=0.
\ee
In general, the linearly independent solutions are given by the Whittaker functions $M_{im,\mu}(-2i\omega/r)$ and $W_{im,\mu}(-2i\omega/r)$ (see e.g.,~\cite{gr}). Below, we write formulas for the generic case $2 \mu \notin \Z$.  The special cases $2 \mu \in \Z$ can be recovered by careful evaluation of the appropriate limits.  The only novel feature is the appearance of a logarithm in the large $r$ expansion for $\mu=0$, a case that one does not expect to arise for massless scalars.

\subsection{The inner product and boundary conditions}
\label{phiBC}

A central object in the study of scalar fields is
the Klein-Gordon current
\be
j_a(\Phi_1 , \Phi_2)=-i ( \Phi_1 \partial_a \Phi_2 ^*- \Phi_2 ^*\partial_a \Phi_1).
\ee
Conserved charges are readily calculated from this current, and the associated Klein-Gordon norm
\be
\label{KGN}
\Omega_\Sigma( \Phi_1, \Phi_2)=\int_\Sigma \sqrt{g_\Sigma}~j^at_a~d\theta d\phi dr,
\ee
plays a key role in quantizing the field and in constructing the classical phase space.  In (\ref{KGN}), $t_a$ is the unit one-form normal to the spatial hypersurface $\Sigma$ and $g_\Sigma$ is the induced metric on $\Sigma$.   At least in anti-de Sitter space, choices of boundary conditions under which $\Omega_\Sigma$ is finite and conserved are closely related to those which define a self-adjoint Hamiltonian, and thus which have a well-defined Cauchy problem in the sense of $L^2$ functions.   This can be seen by comparing the analogue of our discussion below with e.g. \cite{Ishibashi:2004wx}.

We therefore require the norm (\ref{KGN}) to be finite and conserved.  Let us first examine the normalizeability of our asymptotic solutions $r^{- \frac{1}{2} + \mu}$ at large $r$.  Modes with $\Re[\mu] <0$ (``fast fall-off modes'') are always normalizeable at infinity, as are in fact all modes with $\Re[\mu]<1/2$.  In particular, this includes all modes with imaginary $\mu$.

Since we work in Poincar\'e coordinates, we must also consider normalizeability at the horizon.  At nonzero frequency, the horizon is an irregular singular point and any solution behaves as a superposition of modes with $R\sim\exp(\pm i \omega / r)$. For real frequency, this is always delta-function normalizeable at $r=0$. But this is no longer the case for complex frequencies.   In the following discussion we will consider frequencies in the upper half plane, $\omega = | \omega |e^{i\gamma}$ where $0\le\gamma\le\pi$, so that normalizeability requires the $\exp(+i \omega/r)$ behavior near the horizon for $\gamma \neq 0, \pi$. The situation for the lower half plane is analogous, with appropriate changes of signs.   In terms of the Whittaker functions, for $\Im[\omega] > 0$ the solution normalizeable on the Poincar\'e horizon is
\be R(r)\propto W_{im,\mu}(-2i \omega/r).
\label{RPhilarger}
\ee
At large radius, (\ref{RPhilarger}) is $R\sim A r^{-1/2+\mu}+B r^{-1/2-\mu}$ with

\be
\label{PhiAB}
\frac{A}{B}=\frac{e^{i\mu(\pi-2\gamma)}\Gamma(2\mu)\Gamma(\frac{1}{2}-\mu-im)}{|2\omega|^{2\mu}\Gamma(-2\mu)\Gamma(\frac{1}{2}+\mu-im)}.
\ee

It is illustrative to consider boundary conditions which act separately on each harmonic on the squashed sphere; i.e., which leave modes with different values of $\ell,m$ uncoupled.  Since our theory is a toy model for linearized gravity, we consider linear boundary conditions: $A_{\ell,m}=\alpha_{\ell,m} B_{\ell,m}$ for each $(\ell,m)$.  To ensure a good classical phase space and the right setting for quantization, we must impose boundary conditions that conserve the Klein-Gordon norm; i.e.,
the flux through a constant $r$ surface must vanish at each time as $r\rightarrow\infty$:
\be
\label{KGF}
\F(\Phi_1, \Phi_2)=\int_{r=\infty}\sqrt{- \gamma}~j^ar_a~d\theta d\phi =0,
\ee
where $r_a$ is a unit one-form normal and $\gamma_{ab}$ is the induced metric  on surfaces of constant $r$.  Noting that modes with different $(\ell,m)$ are orthogonal,
let us consider the flux for a solution asymptotically  of the form $R\sim A r^{-1/2+\mu}+B r^{-1/2-\mu}$.  We have
\begin{eqnarray}
\F(\Phi, \Phi) &\sim& (\mu^*-\mu) |A|^2 r^{\mu +\mu^*} + (\mu -\mu^*) |B|^2 r^{-\mu -\mu^*} \\
&&\qquad\quad + \ (\mu+\mu^*)B A^* r^{\mu^*-\mu}-(\mu+\mu^*)A B^* r^{\mu-\mu^*}+\ldots  \,.
\end{eqnarray}
For real $\mu$ (power law modes), the condition $\F = 0$ becomes $|B|^2 (\alpha - \alpha^*) = 0$, which is solved for $\alpha_{\ell,m}\in \mathbb{R}$. For pure imaginary $\mu$ (oscillatory modes), the condition $\F = 0$ becomes $|B|^2(1-|\alpha|^2) = 0$, which is solved for
$\alpha_{\ell,m}=e^{i\beta_{\ell,m}},~\beta_{\ell,m} \in\mathbb{R}$.

For power law modes, (\ref{PhiAB}) must be real.  This requires
\be
\label{plgamma}
e^{2i\mu(\pi-2\gamma)}=\frac{\cosh\pi(m+i\mu)}{\cosh\pi(m-i\mu)},
\ee
and restricts frequencies with $\Im[\omega] > 0$ to satisfy
\be
\label{123}
\alpha=-|2\omega|^{-2\mu}\frac{\Gamma(1+2\mu)}{\Gamma(1-2\mu)}\left|\frac{\Gamma(\frac{1}{2}+im+\mu)}{\Gamma(\frac{1}{2}+im-\mu)}\right|.
\ee
It is clear that (\ref{plgamma}) admits at most one solution for $\pi > \gamma > 0$, and numerical investigations show that indeed a solution always exists for $|\mu| < 1/2$.  On the other hand,
there are no solutions to (\ref{123}) for a certain sign of $\alpha \in \mathbb{R}$.  In that case, no frequencies are allowed with $\Im[\omega] > 0$, and one may check that the same is true for $\Im[\omega] < 0$.  All solutions have $\Im[\omega]=0$ and are stable.  For the other sign of $\alpha$,
there is a single unstable mode.  Perhaps the most natural choice for $\alpha$ is the so-called generalized Dirichlet boundary condition, which corresponds to the borderline case $\alpha =0$ (equivalently, $A=0$), where we now take $\mu > 0$.  Since the right-hand-side of (\ref{123}) cannot vanish, this boundary condition again allows only real frequencies.

Now consider the oscillatory modes $\mu=ik$ with $k \in \mathbb{R}$. Since $|\alpha| =1$, (\ref{PhiAB}) requires
\be
\label{ogamma}
\gamma=\frac{\pi}{2}+\frac{1}{4k}\ln\left( \frac{\cosh\pi(k+m)}{\cosh\pi(k-m)}\right)
\ee
for $\Im[\omega] > 0$. This is in fact a monotonically increasing function of $m$ with $\lim_{m\rightarrow\pm\infty}=\frac{\pi}{2}\pm\frac{\pi}{2}$, so that a solution in the desired range exists for all $k,m$.

The particular value of the phase $\beta$ determines the magnitudes of the allowed frequencies through

\be
\exp(i\beta)=-|2\omega|^{-2ik}\frac{\Gamma(1+2ik)}{\Gamma(1-2ik)}\sqrt{ \frac{\Gamma(\frac{1}{2}-ik-im)\Gamma(\frac{1}{2}-ik+im)}{\Gamma(\frac{1}{2}+ik-im)\Gamma(\frac{1}{2}+ik+im)} }.
\label{unstablescalarphase}\ee
Since the shift $| \omega |\rightarrow | \omega |e^{\pi/k}$ is a symmetry of (\ref{unstablescalarphase}), for any choice of $\beta$ there are an infinite number of unstable modes logarithmically distributed in frequency along the ray $\arg(\omega)=\gamma$.

\subsection{The linearization-stability constraints}
\label{LSS}

Taking the linear Klein-Gordon field as a toy model of linearized gravity, it is
interesting to seek solutions for which all $SL(2,\R) \times U(1)$ charges vanish.
This would be the analogue of enforcing the linearization-stability constraints in
the gravitational theory.

It is enlightening to begin by discussing the energy, $Q^{lin}_{\eta_1}$.   It is
clear that  the constraint $Q^{lin}_{\eta_1} =0$ admits a large space of solutions.
Consider for example any complex-frequency mode that satisfies any of the
time-independent boundary conditions described in section \ref{phiBC}.  On general
grounds, time-independent boundary conditions conserving the Klein-Gordon norm also
conserve energy.  But since the mode has complex frequency, its
charge $Q^{lin}_{\eta_1}$ must increase (or decrease) exponentially in time.   Hence, $Q^{lin}_{\eta_1} =0$ for such modes.  Furthermore,
linear combinations of growing and decaying modes can have either sign of the
energy, so there is much freedom in solving this constraint.

Suppose that we now take the surface $\Sigma$ on which the charges are evaluated to
be just $t=0$.  Since the dilatation $\eta_0$ is spacelike at $t=0$ (see Figure
\ref{AdS2}), the associated charge $Q^{lin}_{\eta_0}$ is effectively a momentum and
one can find zero-energy solutions having either sign of this charge.  It is then
straightforward to find a linear combination $C\Phi_1 + D \Phi_2$ of two zero-energy
modes having different values of $\ell,m$ for which $Q^{lin}_{\eta_0}$ vanishes at
$t=0$.  Using $\pounds_{\eta_1} \Phi_j = - i \omega \Phi_j$ , the $SL(2,\R)$ algebra, and
the fact that modes with distinct $(\ell,m)$ are orthogonal under $\Omega_\Sigma$, it
then follows that $Q^{lin}_{\eta_{-1}}$ also vanishes for $C\Phi_1 + D \Phi_2$.  The
final constraint $Q^{lin}_{\xi_0} = 0$ can be satisfied by  combining two such
solutions with opposite signs of $m$.  As a result, the full set of
linearization-stability constraints at $t=0$ admits a large space of simultaneous
solutions.

However, the charges $Q^{lin}_{\eta_0}$ and
$Q^{lin}_{\eta_{-1}}$ are not conserved by the boundary conditions of section
\ref{phiBC}.  The problem can be stated in simple physical terms.  To do so, recall
from \cite{Bardeen:1999px} that timelike geodesics (i.e., particle trajectories) in
the extreme Kerr throat can reach the boundary in finite coordinate time $t$.
Recall also that such geodesics are associated with oscillatory modes via the WKB
approximation.   As a result, the $r = \infty$ boundary acts like a wall at finite
distance with respect to such modes.  Any boundary condition that conserves
the Klein-Gordon norm effectively causes particles to reflect off of this boundary,
perhaps with some phase shift. Now, we noted above that $\eta_0$ is spacelike at
$t=0$, and that the associated $Q^{lin}_{\eta_0}$ is therefore a sort of momentum.
But it is clear that reflections off of a finite-distance wall cannot conserve
momentum: particles incident on the wall arrive from the bulk (say, to the left of the wall) and so necessarily have one sign of momentum. Particles leaving the wall must return to the bulk and so necessarily have the opposite sign.  In much
the same way, direct calculation shows that the flux of $Q^{lin}_{\eta_0}$ due to
the oscillatory modes through the surface $r = \infty$ is positive definite (for real solutions) at $t=0$.  At general $t$ (and for general complex solutions), one may construct a positive-definite combination of this flux $(\F_{\eta_0})$ and the energy flux $(\F_{\eta_1})$.       The
same is true for $Q^{lin}_{\eta_{-1}}$:
\begin{eqnarray}
\label{c3}
(\F_{\eta_0} - t\F_{\eta_1})^{\rm oscillatory}&\propto& \sum_{m,\ell \,{\rm with}\, K_{m,\ell} < K_{crit}
}(|A_{m,\ell}|^2 + |B_{m,\ell}|^2)k^2\,,\\
(\F_{\eta_-1} - t^2 \F_{\eta_1})^{\rm oscillatory}&\propto& \sum_{m,\ell
\,{\rm with} \,K_{m,\ell} < K_{crit} }(|A_{m,\ell}|^2 + |B_{m,\ell}|^2) k^2 t\,.
\end{eqnarray}

For any Dirichlet-type boundary conditions, the flux from power law modes
vanishes.  As a result, for such boundary conditions the only solution satisfying
the constraints at all times is $\Phi =0$.
It is worth noting, however, that this amounts to a failure of the Cauchy problem
for such boundary conditions:  We have valid initial data which satisfies all
constraints and boundary conditions at $t=0$.  However, there is no evolution of
this data which satisfies the boundary conditions for all $t$.

The reader may wonder whether some more general boundary condition would allow
additional solutions.  For example, one might ask if a nonlinear boundary condition
for each mode could preserve the $SL(2,\mathbb{R})$ symmetries.   However, this
would require $A_{\ell,m}\propto B_{\ell,m}^{\frac{1-2ik}{1+2ik}}$,  which does not conserve
the appropriate flux\footnote{The analysis of non-linear boundary conditions is
similar to that of linear boundary conditions.  Symplectic flux would be conserved
for power law modes only if small variations $\delta A_{\ell,m}$ and  $\delta
B_{\ell,m}$ are related by a phase. Even for real $B_{\ell,m}$, this is true for  $A_{\ell,m}\propto B_{\ell,m}^{\frac{1-2ik}{1+2ik}}$ only for $k = 0$, a special case which requires separate analysis due to the appearance of a logarithmic mode.}.

Returning to linear boundary conditions, one might also try to allow suitable linear
combinations of fast- and slow-fall-off solutions with $\mu < 1/2$.  Indeed, the
fluxes from power law modes satisfy
\begin{eqnarray}
\label{c4}
(\F_{\eta_0} - t\F_{\eta_1})^{\rm power\ law}&\propto& \sum_{m,\ell \, {\rm with}\, K_{m,\ell} > K_{crit} }
\alpha_{m,\ell} |A_{m,\ell}|^2 k^2 \,, \\
(\F_{\eta_-1} - t^2\F_{\eta_1})^{\rm power\ law}&\propto& \sum_{m,\ell \,
{\rm with} \,K_{m,\ell} > K_{crit} }\alpha_{m,\ell} |A_{m,\ell}|^2 k^2 t,
\end{eqnarray}
and in particular are negative for modes with $\alpha_{m,\ell} < 0$.   While we have
not analyzed this possibility in full detail, it is difficult to imagine a boundary
condition which achieves this while simultaneously conserving Klein-Gordon flux.  In
particular, while one can tune the magnitudes of the frequencies $|\omega|$ of the unstable modes (in
both power law and oscillatory cases) through a choice of boundary condition, at least
with the boundary conditions of section \ref{phiBC} the phase of $\omega$ is a
fixed, complicated function of $m, \mu$.  It is therefore difficult to cancel the
flux due to an unstable oscillatory mode against the flux from an unstable
power law mode for all times.  This is an interesting departure from the analogy
with AdS${}_2$ in the presence of scalars both just above and just below the BF bound.  In that case,  the unstable frequencies were always purely imaginary and one could easily find boundary
conditions which admit solutions where all charges vanish for all time.

\setcounter{equation}{0}

\section{Linearized gravitational perturbations}
\label{grav}

We now analyze linearized gravitational waves in the NHEK background, adapting technology developed by Teukolsky \cite{Teukolsky:1973ha} for perturbations of asymptotically flat Kerr black holes.   While a straightforward analysis of linearized gravity leads to separable, decoupled equations for highly symmetric backgrounds like Schwarzschild \cite{nonrotating}, the same is not true for Kerr.  Instead, a more subtle approach is required. Teukolsky showed that the essential field equations for spin $0,\pm 1/2, \pm 1, \pm 2$ decouple using a Newman-Penrose approach, and that separating variables then leads to ordinary differential equations as usual. The Newman-Penrose null tetrad formalism \cite{Newman:1961qr} is briefly reviewed in Appendix A.

The main result of this section is that the behavior of linearized gravitons (up to linearized diffeomorphisms) is directly analogous to that found in section \ref{scalars} for linear scalar fields: Modes with large $m$ and small $K$ oscillate near infinity.  For such modes, the flux of the linearized charge $Q^{lin}_{\eta_0}$ is positive definite.  It is difficult to balance this positive flux against a negative flux from power law modes, and impossible for the analogue of generalized Dirichlet boundary conditions.    We therefore expect that, when the full set of non-linear couplings are taken into account, the only linearized solutions satisfying all linearization-stability constraints will be linearized diffeomorphisms.

\subsection{Spin-$s$ Teukolsky equations}
\label{spinseqs}

In the Newman-Penrose formalism, the gravitational field is described in part by the scalars $\psi_0,\psi_1, \psi_2,\psi_3,\psi_4$, which are certain components of the Weyl tensor.  For the Kerr background, it turns out that all of these scalars vanish except $\psi_2$; as noted in \cite{wald:1453}, the fact that many background quantities vanish is a promising signal that the perturbation analysis in these new variables will simplify.  Indeed, both Schwarzschild and Kerr are classified as type $D$ spacetimes, and so within the Newman-Penrose framework, the rotating and non-rotating cases are actually quite similar.

For the case of interest, spin $\pm2$, solutions to the Teukolsky equation only give us the form of the Weyl scalars $\psi_0, \psi_4$, and one may question if this is enough to fully specify the gravitational perturbations.  In fact, it was shown by Wald that physical graviton fluctuations of Kerr are encoded entirely in either $\psi_0$ or $\psi_4$, which are invariant under linearized diffeomorphisms. The only perturbations with $\psi_0=\psi_4=0$ are {\it i\,}) deformations that either change the mass or angular momentum of the Kerr black hole, {\it ii\,}) a linearized deformation towards the rotating C-metric, or {\it iii\,}) a linearized deformation adding NUT charge \cite{wald:1453}.

Solutions of the Teukolsky equation in the NHEK geometry will therefore provide all nontrivial perturbations up to diffeomorphisms; i.e., up to the possible presence of boundary gravitons.   To carefully analyze valid boundary conditions, it is also useful to have a method of translating from the Weyl scalars to actual metric fluctuations. Fortunately, the details of this ``inversion problem'' have been worked out for the full Kerr geometry in \cite{Chrzanowski:1975wv}, so one simply has to take the appropriate limit of this procedure to find the NHEK metric fluctuations.

We now briefly review the Teukolsky equations \cite{Teukolsky:1973ha} for the Kerr background in Boyer-Lindquist coordinates $(\tilde t, \tilde r, \theta, \tilde \phi)$. The Teukolsky equations for spin $s$ are differential equations for certain scalar quantities $\psi^{(s)}$.  For gravitational perturbations in particular, we have $\psi^{(2)} = \psi_0$ and $\psi^{(-2)} = \rho^{-4} \psi_4$, where $\rho$ is the spin coefficient defined in appendix \ref{NP}.  Separating variables as $\psi^{(s)} =e^{-i\omegah \th+im\phih}R_s(\rh)S_s(\theta)$, the spin-$s$ Teukolsky equations are
\be
\Delta \frac{d^2R_s}{d\rh^2}+2(s+1)(\rh-M)\frac{dR_s}{d\rh}+\left(\frac{C^2-2is(\rh-M)C}{\Delta}+4is\omegah \rh-\Lambda\right)R_s=0,\label{radialspinsfull}
\ee
\begin{eqnarray}
\nonumber
\frac{1}{\sin\theta}\frac{d}{d\theta}\left(\sin\theta\frac{dS_s}{d\theta}\right)
+\left( a^2\omegah^2\cos^2\theta  -\frac{m^2}{\sin^2\theta}
-2a\omegah s\cos\theta \right.&& \\
&&\hspace{-6.5cm}\left.-\frac{2ms\cos\theta}{\sin^2\theta}-s^2\csc^2\theta-\frac{m^2}{4}+K\right)S_s=0 \,,
\end{eqnarray}
where
\begin{equation}
\Delta=\rh^2-2M\rh+a^2\,, \qquad C=(\rh^2+a^2)\omegah-am
\end{equation}
and
\begin{equation}
\Lambda=K-m^2/4+a^2\omegah^2-2am\omegah-s(s+1) \,.
\end{equation}
Our eigenvalue $K$ is related to the Teukolsky eigenvalue $A$ by $A=K-m^2/4-s(s+1)$, so that in the near horizon limit, we recover the $s=0$ equation given in \cite{Bardeen:1999px}. It is worth noting that our constant $K$ is not at all related to Teukolsky's $K$, which has radial dependence.

To find the near-horizon form of these equations for extreme Kerr, we apply the change of variables (\ref{nhekscaling}) 
and define a shifted frequency $\omega$ through $\omegah=\lambda\omega+m/(2M)$.  For convenience, we set the length scale $r_0$ to unity, $r_0^2=2M^2=1$. As in the scalar case, this can be done with the simple rescaling of $r\rightarrow r_0 r,~\omega\rightarrow \omega/r_0 $.  Taking the limit $\lambda\rightarrow 0$, the Teukolsky equations become
\be
r^2R_s''+2r(1+s)R_s'+\left(2m^2-K+s(1+s)+\frac{2\omega(m-is)}{r}+\frac{\omega^2}{r^2}\right)R_s=0
\label{spinsradial}\ee
\begin{eqnarray}
\nonumber
\frac{1}{\sin\theta}\frac{d}{d\theta}\left(\sin\theta\frac{dS_s}{d\theta}\right)
-\left(\frac{m^2+s^2+2 m s \cos\theta}{\sin^2\theta}\right)S_s&&\\
&&\hspace{-6cm}+\left(\frac{m^2}{4}\cos^2\theta-m s \cos\theta\right)S_s+\left(K-\frac{m^2}{4}\right)S_s=0\,.
\label{spinsspherical}
\end{eqnarray}

The radial equation (\ref{spinsradial}) is simply a deformation of the scalar wave equation in Poincar\'e coordinates, and the exact solutions are discussed in the next section. The large $r$ behavior is easily seen to be
\be
R_s\approx r^\Delta,~\Delta=-\frac{1}{2}-s + \mu\, ,
\ee
where again $\mu^2 = K -  2m^2 + 1/4$.
For $K<K_{crit}^s=2m^2-1/4$, one finds complex exponents in parallel with the oscillatory scalar modes and with scalars in AdS below the Breitenlohner-Freedman bound. 

The angular wave equation is a deformation of the equation describing spin-weighted spherical harmonics ${}_sY^m_\ell=Y(\theta) e^{i m \phi}$, where
\be
\frac{1}{\sin\theta}\frac{d}{d\theta}\left(\sin\theta\frac{dY}{d\theta}\right)
-\left(\frac{m^2+s^2+2 m s \cos\theta}{\sin^2\theta}\right)Y=- \ell (\ell+1) Y,
\ee
and the eigenvalues take the familiar form
\be
\quad \ell=|s|,|s|+1,\dots, \quad -\ell\le m \le +\ell\,.
\ee
As a result, for axisymmetric perturbations (i.e., $m = 0$), the exact solutions to the angular Teukolsky equation are just ${}_sY^0_\ell$ with $K = \ell(\ell+1)$.  While a full analytic treatment is not available for $m \neq 0$, as in section \ref{scalars} we can consider the regime $\ell=m\gg1$ with modes localized near the equator.  We again find the eigenvalues $K_{\ell=|m|\gg1}=5/4 m^2+\Op(m)$, which correspond to $\mu=\frac{\sqrt{3}}{2}im+\Op(1)$.

\subsection{Solving the radial wave equation}
\label{radsolve}
To solve the radial equation, we define a new function $M(r) = r^{s} R(r)$ and  make the change of variable $z=-2i\omega/r$.  The wave equation then takes the form
\be
M''(z)+\left(-\frac{1}{4}+\frac{im+s}{z}+\frac{1/4-\mu^2}{z^2} \right)M(z)=0\,.
\ee
In general, the linearly independent solutions are given by the Whittaker functions $M_{im+s,\mu}(-2i\omega/r)$ and $W_{im+s,\mu}(-2i\omega/r)$.  Below we restrict to the generic case $2 \mu \notin \Z$.   Similar results hold for the special cases $2 \mu \in \Z$ so long as $\mu \neq 0$.
We assume below that the complicated spectrum of $K$ on the squashed sphere forbids the logarithmic case $\mu=0$.  

The condition that modes with $\Im[\mu] > 0$  be normalizeable on the Poincar\'e horizon requires 
\be
R(r) \propto r^{-s} W_{im+s,\mu}(-2i\omega/r), \ \ \
\Im [\omega] > 0.  \ee
As $r\to \infty$, this solution behaves as $A r^{-1/2-s+\mu}+B r^{-1/2-s-\mu}$, with the ratio of coefficients
\be\label{teukratio}
\frac{A}{B}=-\frac{\Gamma(1+2\mu)\Gamma(\frac{1}{2}-im-s-\mu)}{\Gamma(1-2\mu)\Gamma(\frac{1}{2}-im-s+\mu)}(2|\omega|)^{-2i\mu}e^{-2\mu(\gamma-\pi/2)},
\ee
where we have written the frequency as $\omega=|\omega|e^{i\gamma}$ for some phase $\pi > \gamma > 0$.  Note that the restriction on the range of $\gamma$ means that, for unstable modes, $A/B$ can take values only in half of the complex plane.  An analogous condition holds for $\Im[\omega] <0$.    There is no such requirement for real frequencies, so that the cases $\gamma=0, \pi$ are also allowed. See e.g. \cite{Detweiler:1980gk, Hod:2008zz}
for similar discussions in the context of asymptotically flat Kerr black
holes.

\subsection{Construction of the metric perturbation}
\label{inverse}
In order to analyze boundary conditions for asymptotically NHEK solutions, we would like to know the large $r$ behavior of the actual metric perturbations $h_{ab}$.  Due to certain special properties of the Teukolsky differential operators, it turns out that one can solve this ``inverse'' problem simply by
taking various derivatives of the solutions to the Teukolsky equations.   This was  first argued in \cite{Chrzanowski:1975wv} using Green's functions for the gravitational perturbations, and then more generally using self-adjointness properties of the equations in \cite{Wald:1978vm}.   The key result for constructing metric perturbations about the Kerr spacetime is \cite{Chrzanowski:1975wv}
\begin{eqnarray}
\label{hrecon}
\nonumber h_{ab}&=&\{-l_al_b(\delta^*+\alpha+3\beta^*-\tau^*)(\delta^*+4\beta^*+3\tau^*)-m^*_am^*_b(D-\rho^*)(D+3\rho^*) \\
\nonumber
&&+l_{(a}m^*_{b)}[(D+\rho-\rho^*)(\delta^*+4\beta^*+3\tau^*)+(\delta^*+3\beta^*-\alpha-\pi-\tau^*)(D+3\rho^*)]\}\\
&&\times R_{-2}(\rh)S_{2}(\theta)e^{im\phih-i\omega \th}.
\end{eqnarray}
This gives a solution to the linearized Einstein equations in an ``ingoing'' gauge, satisfying $h_{ab} l^b = 0 = h^a{}_a$.
Note that this relation involves the solutions to the $s = -2$ radial equation and the $s = +2$ angular equation.  We denote the metric perturbation built from $M_{im+s, \mu}$ as $h_{ab}^{( \mu)}$ where $\mu$ can have either sign and be either real or imaginary.

To compute this metric perturbation in the near-horizon spacetime, we use results for the NHEK tetrad and spin coefficients given in appendix \ref{NP}, combined with the radial solutions discussed in the previous section.  For large $r$, we find that the perturbation behaves as
\be
\label{Tpert}
h^{(\mu)}_{ab}=
\left(\begin{array}{cccc}h_{tt}=\Op(r^{3/2 + \mu}) & h_{tr}=\Op(r^{-1/2+ \mu}) & h_{t\theta}=\Op(r^{1/2+ \mu})& h_{t\phi}=\Op(r^{1/2+ \mu}) \\
& h_{rr}=\Op(r^{-5/2+ \mu} )& h_{r\theta}=\Op(r^{-3/2+ \mu}) & h_{r\phi}=\Op(r^{-3/2+ \mu}) \\
&  & h_{\theta \theta}=\Op(r^{-1/2+ \mu} )& h_{\theta\phi}=\Op(r^{-1/2+ \mu}) \\
& &  & h_{\phi \phi}=\Op(r^{-1/2+ \mu})
\end{array}\right).
\ee
For the special range $\Re[\mu] < 1/2$ (see below), the components $h_{ab}^{(\mu)}$ are each subleading in $r$ to the corresponding NHEK background metric components (for the non-zero components of (\ref{nhekmetric})).

Since the solutions come in pairs with values $\pm \mu$, the natural analogue of Dirichlet boundary conditions would be to forbid all modes with $\Re[ \mu] > 0$;  i.e., one would require the full perturbation to satisfy (\ref{Tpert}) with $\mu =0$.  We term these ``Teukolsky-Dirichlet boundary conditions."  However, as in our discussion of scalar fields, one still requires some additional boundary condition for modes with imaginary $\mu$.

\subsection{The inner product and boundary conditions}
\label{Tbcs}

In this section, we investigate valid  boundary conditions for linearized metric perturbations in the NHEK geometry.  We must impose boundary conditions so that the inner product (technically, the symplectic structure) is both finite and conserved.  This is the key condition ensuring that our theory has a well-defined phase space.  For scalar fields, the symplectic structure is simply the familiar Klein-Gordon inner product which we discussed in section \ref{scalars}.    Following  \cite{Guica:2008mu}, we will adopt the covariant phase space  formalism of \cite{Barnich:2001jy,Barnich:2007bf}, in which the symplectic current for metric perturbations in Einstein-Hilbert gravity takes the form
\be
\label{symplectic}
\omega_{EH}^a[\delta_1g,\delta_2g]=-P^{abcdef}\left( \delta_2g_{cd} \nabla_b \delta_1 g_{ef}-(1\leftrightarrow 2)\right),
\ee
where
\begin{eqnarray}
\nonumber
P^{abcdef}&=&\frac{1}{32 \pi G} \left(g^{ab} g^{e(c}g^{d)f}+g^{cd} g^{a(e}g^{f)b}+g^{ef} g^{a(c}g^{d)b} \right. \\
&& \left. \qquad\qquad-g^{ab} g^{cd}g^{ef}-g^{a(e} g^{f)(c}g^{d)b}-g^{a(c} g^{d)(e}g^{f)b}\right)\,.
\end{eqnarray}
Note that this leads to a symplectic structure that differs from \cite{Wald:1999wa} by a boundary term. It might also be interesting to explore the addition of further boundary ``counter-terms'' in analogy with those studied in \cite{Compere:2008us} for AdS, but we will not do so here.

Let $\Sigma$ be a constant-time hypersurface with unit normal $t^a$.  Then, given a background metric $g$ and two linearized perturbations $\delta_1 g, \delta_2 g$, the symplectic structure associated with $\Sigma$ is
\begin{equation}
\label{gSymp}
\Omega_\Sigma(g; \delta_1 g, \delta_2 g) = \int_\Sigma d\theta d\phi dr \sqrt{g_\Sigma} \,t_a \omega^a(g; \delta_1 g, \delta_2 g) \,.
\end{equation}
One has $\nabla_a\omega^a=0$
for perturbations satisfying the linearized equations of motion.  As usual, normalizeability at the horizon implies that the symplectic flux through the horizon vanishes.  Hence, $\Omega_\Sigma$ will be conserved if the flux $\F$ through the boundary at $r\to\infty$ vanishes at each time.  Letting $r^a$ denote the unit normal and $\gamma_{ab}$ the induced metric on constant $r$ surfaces,  this flux is
\begin{equation}
\F(g; \delta_1 g, \delta_2 g) = \int_{r = \infty} d\theta d\phi  \sqrt{-\gamma} \,r_a \omega^a (g; \delta_1 g, \delta_2 g)  \,.
\end{equation}

Using (\ref{gSymp}), one finds that the $h^{(\mu)}$ Teukolsky metric perturbations \eqref{Tpert} are normalizeable only when $\Re[\mu] < 1/2.$ Note that this range includes all the oscillatory modes ($K < K_{crit}$), which are characterized by a purely imaginary $\mu$.  For real $\mu$ with $|\mu| \geq 1/2$, normalizeability requires that the slow fall-off mode ($\mu > 0$) be fixed (i.e., a Dirichlet-type boundary condition).

For real $\mu$, $0 \le |\mu| <1/2$, both linearly independent $\pm \mu$ modes are normalizeable and we have a choice of boundary conditions at infinity. This is very much analogous to the mass range near the Breitenlohner-Freedman bound for scalar fields in AdS.   A natural choice is to allow only the ``fast fall-off mode" ($\mu <  0$) for real $\mu$; i.e., Teukolsky-Dirichlet boundary conditions.   For imaginary $\mu$, both modes are normalizeable and there is no distinguished boundary condition.  We once again expect instabilities for this case.

It is illustrative to consider boundary conditions which act separately on each harmonic on the squashed sphere; i.e., which leave modes with different values of $\ell,m$ uncoupled.  Since we work in the linearized theory, we consider linear boundary conditions : $A_{\ell,m}=\alpha_{\ell,m} B_{\ell,m}$.
To determine the allowed coefficients $\alpha_{\ell,m}$, consider the flux due to the symplectic product between an oscillatory mode with $\mu = i k, k\in  \R$ and the complex conjugate of an oscillatory mode with $\mu = i k', k'\in  \R$ (it is sufficient to consider modes with the same $m$ since the flux clearly vanishes otherwise).  We have
\bea
\nonumber
&&\!\!\!\!\!\!\!\!\!\!\!\!\!\!\!\!\F(A_1 h^{(ik)}+B_1 h^{(-ik)},A_2^* h^{(ik')*}+B_2^* h^{(-ik')*})
\\ \nonumber &&= A_1 B_2^* \F(h^{(ik)},h^{(-ik')*})
+B_1 A_2^* \F(h^{(-ik)},h^{(ik')*})+A_1 A_2^* \F(h^{(ik)},h^{(ik')*}) \\
&&\quad+B_1 B_2^* \F(h^{(-ik)},h^{(-ik')*}).
\eea
Explicit computation shows that this expression always vanishes unless $k = k'$ due to orthogonality of the angular functions $S^{(\mu)}(\theta)$.  In this case, the first two terms are each zero due to the anti-symmetry of the symplectic structure.  We must then choose the boundary condition for this mode so that the last two terms cancel. This fixes the relevant value of $\alpha = A_1/B_1 = A_2/B_2$.  After some lengthy calculations, we find
\be
\label{alpha2}
|\alpha|^2=\frac{9+40(k+m)^2+16(k+m)^4}{9+40(k-m)^2+16(k-m)^4} \,.
\ee
We see that the allowed boundary conditions of the above form are parametrized by a choice of phase for each mode.

Once an $\alpha$ satisfying (\ref{alpha2}) is chosen, the spectrum of frequencies $\omega = |\omega| e^{i \gamma}$ can be determined.   Since (\ref{alpha2}) is independent of $\omega$, all  real frequencies ($\gamma=0$) are allowed.  Recall, however, that for complex frequencies we have an additional constraint \eqref{teukratio} on $|\alpha|^2$ from regularity of the Teukolsky scalar at the horizon. Using that the radial Teukolsky function solves the $s=-2$ radial equation, for $\Im[\omega] >0$ this constraint yields
\be
\label{alphaT}
|\alpha|^2=\left|\frac{A}{B}\right|^2=e^{2k(2\gamma-\pi)}
\left(\frac{9+40(k+m)^2+16(k+m)^4}{9+40(k-m)^2+16(k-m)^4}\right)\,\frac{\cosh(\pi(k-m))}{\cosh(\pi(k+m))} \,.
\ee
Remarkably, the complicated ratio of polynomials in $k,m$ cancels when one compares (\ref{alphaT}) with (\ref{alpha2}).  What remains is a restriction on the phase $\gamma$ of the complex frequency $\omega = |\omega| e^{i \gamma}$ which is precisely the same as  in the scalar field case:

\be
\label{gammaT}
\gamma=\frac{\pi}{2}+\frac{1}{4k}\ln\left( \frac{\cosh\pi(k+m)}{\cosh\pi(k-m)}\right).
\ee
Choosing a time-independent phase for $\alpha = A/B$ then leads to a quantization condition on the magnitudes of the complex frequencies $|\omega|$ through the restriction that  \eqref{teukratio}  has the correct phase.  One may also choose a time-dependent phase, though of course this breaks time-translation symmetry so that modes with definite frequency are no longer solutions.

For the power law modes with $\mu, \mu' < 1/2$, one can perform a similar calculation of the flux
$\F(A_1 h^{(\mu)}+B_1 h^{(-\mu)},A_2^* h^{(\mu')*}+B_2^* h^{(-\mu')*})$.
Once again, $\F = 0$ when $\mu \neq \mu'$.  For $\mu = \mu'$,  imposing a boundary condition $A_1 = \alpha B_1, A_2 = \alpha B_2$ as above now restricts the phase (rather than the magnitude) of $\alpha$:
\begin{equation}
\label{subcritT}
\frac{\alpha}{\alpha^*} = \frac{9+40(m-i \mu)^2+16(m-i \mu)^4}{9+40(m+i \mu)^2+16(m+i \mu)^4}   \,.
\end{equation}
Note that two opposite phases are allowed by (\ref{subcritT}).  But it was noted below (\ref{teukratio}) that unstable solutions exist only when $\alpha$ lies in some particular half of the complex plane.  Thus, just as in the scalar case,  one choice of phase leads only to stable modes, while the other again leads to a single unstable mode with $\gamma$ given by (\ref{plgamma}).  Perhaps the most natural choice for $\alpha$ is the Teukolsky-Dirichlet boundary condition $\alpha=0$ (equivalently, $A=0$), where we now take $\mu > 0$.  
While (\ref{subcritT}) degenerates for this case, one may note that  (\ref{teukratio}) admits no solutions.  As a result, only real frequencies are allowed.

\subsection{Charges and constraints}
\label{Tcharges}

Finally, we discuss the linearized charges of the above solutions in connection with the conjectured linearization-stability constraints.  Recall that the condition that the charges generate $\xi$-translations, and the fact that the symplectic structure is the inverse of the Poisson bracket, imply that the linearized charge associated with an isometry $\xi$ can be written in terms of the symplectic structure about the background $\bar g$ as
\begin{equation}
\label{linTcharge}
Q^{lin}_\xi = \frac{1}{2} \Re \  \Omega_\Sigma(\bar g; \pounds_\xi h , h^*) \,.
\end{equation}

It is enlightening to begin by discussing the energy, $Q^{lin}_{\eta_1}$.  Note that $Q^{lin}_{\eta_1}$ is conserved under any boundary conditions for which i) the symplectic structure $\Omega$ is conserved and ii) the boundary conditions are invariant under $\eta_1$, so that $\pounds_{\eta_1} h$ satisfies the boundary conditions whenever $h$ does.   This is the case for the time-independent boundary conditions discussed in section \ref{Tbcs}.

As a result, it is clear that the linearization-stability constraint $Q^{lin}_{\eta_1} =0$ admits a large space of solutions.  Consider for example any mode with complex frequency.  The charge $Q^{lin}_{\eta_1}$ carried by any such mode must increase (or decrease) exponentially in time.  But it is also conserved.  Hence, $Q^{lin}_{\eta_1} =0$ for such modes.  Furthermore, linear combinations of growing and decaying modes can have either sign of the energy, so there is much freedom is solving this constraint.

Suppose that we now take the surface $\Sigma$ on which the charges are evaluated to be just $t=0$.  Since the dilatation charge $\eta_0$ is spacelike at $t=0$ (see Figure \ref{AdS2}), it is clear that one can find unstable modes of the type discussed above having either sign of the $\eta_0$ charge.  It is then straightforward to find a linear combination $Ch_1 + D h_2$ of two unstable modes having different values of $\ell,m$ for which  $Q^{lin}_{\eta_0}$ vanishes at $t=0$.  Using $\pounds_{\eta_1} h_j = -i \omega h_j$, the $SL(2,\R)$ algebra, and the fact that modes with distinct $(\ell,m)$ are orthogonal under $\Omega_\Sigma$, it follows that $Q^{lin}_{\eta_{-1}}$ also vanishes for $Ch_1 + D h_2$.  The final constraint $Q^{lin}_{\xi_0} = 0$ can then be satisfied by  combining two such solutions with opposite signs of $m$.  As a result, at $t=0$ there is a large space of simultaneous solutions to the linearization-stability constraints\footnote{\label{extend}The use of modes of definite frequency satisfying given boundary conditions was merely a technical crutch in the above argument; the result holds for very general boundary conditions.   Consider for example some strict Dirichlet boundary condition that fixes $h=0$ at a large but finite value $r_{Dir}$ of $r$.  The spectrum will include both stable and unstable modes, allowing solutions to the constraints to be constructed as above.   We can then extend the corresponding initial data to all $r$ by simply taking it to vanish for $r > r_{Dir}$.  The resulting data has a discontinuity in its first $r$-derivative at $r = r_{Dir}$, but nevertheless continues to provide a solution to the $t=0$ linearization-stability constraints.}.  

However, as with our prior discussion of the scalar field, the charges $Q^{lin}_{\eta_0}$ and
$Q^{lin}_{\eta_{-1}}$ are not conserved by the boundary conditions of section \ref{Tbcs}.  The problem is again that conservation of symplectic flux requires fixing the phase of each $\alpha_{\ell,m}$ for the oscillatory modes, but that this breaks the symmetries generated by $\eta_0$ and $\eta_{-1}$.  In fact, before imposing any boundary condition, the fluxes  $\F_{\eta_0} =  \Re \  \F(\bar g; \pounds_{\eta_0} h, h^*)$ and $\F_{\eta_1} = \Re \ \F(\bar g; \pounds_{\eta_1} h, h^*)$ of $Q^{lin}_{\eta_0}$ and the energy $Q^{lin}_{\eta_1}$ can be shown to satisfy

\begin{eqnarray}
\label{c5}
\left( \F_{\eta_0}  - t \F_{\eta_1} \right)^{oscillatory} &=& \sum_{\ell, m \ {\rm with} \ K_{\ell, m} < K_{crit}} \frac{k^2}{4}\left[\left(9+40 (k-m)^2 +16(k-m)^4 \right)|A_{\ell, m}|^2 \right.\nonumber \\
&&\left.\qquad+\left(9+40 (k+m)^2 +16(k+m)^4 \right)|B_{\ell, m}|^2 \right] \,,
\end{eqnarray}
which is positive definite.
In contrast, this combination of fluxes vanishes for power law modes under boundary conditions which, for each $(\ell,m)$, allow only $h^{(\mu)}$ or $h^{(-\mu)}$.  In the range $\mu$ real, $|\mu| < 1/2$, where both modes are normalizeable, linear combinations $Ah^{(\mu)} + Bh^{(-\mu)}$ can have either sign of this flux.

Since we must impose the constraints $Q^{lin}_{\eta_i} =0$ at all times, $\left( \F_{\eta_0}  - t \F_{\eta_1} \right)$ must vanish at each $t$.  From the above discussion, it is clear that this is not possible under Teukolsky-Dirichlet type boundary conditions (\eqref{Tpert} with $\mu=0$), as these forbid the slow fall-off solutions $h^{(|\mu|)}$ for power law modes.  One might have thought that this was the most natural possible boundary condition.  However, we now see that flux conservation for $Q^{lin}_{\eta_{0}}$ forces all oscillatory modes to vanish, and that the energy constraint $Q^{lin}_{\eta_{1}} =0$ then forces all power law modes to vanish (since they are stable, they carry only positive energy).  As a result, Teukolsky-Dirichlet boundary conditions allow only the trivial solution $h=0$.

The reader may wonder whether some more general boundary condition would allow additional solutions. In particular, one might try to allow suitable linear combinations $Ah^{(\mu)} + Bh^{(-\mu)}$ of power law modes for $\mu < 1/2$, hoping to cancel their negative $\eta_0$-flux against the positive $\eta_0$-flux from oscillatory modes.  While we have not analyzed this possibility in full detail, it is difficult to imagine a boundary condition which achieves this while simultaneously conserving symplectic flux.    In particular, while one can tune the magnitudes of the frequencies $|\omega|$ of unstable modes (in both power-law and oscillatory cases) through a choice of boundary condition, at least
with the boundary conditions of section \ref{Tbcs} the phase of $\omega$ is a
fixed, complicated function of $m, \mu$.  It is therefore difficult to cancel the
flux due to an unstable oscillatory mode against the flux from an unstable
power law mode for all times.  

\setcounter{equation}{0}

\section{Discussion}
\label{disc}

We have argued that dynamics in the extreme Kerr throat is highly constrained.  We found that scaling limits of non-extreme Kerr black holes also yield the NHEK geometry, but in coordinates with a finite temperature horizon.  We then proved that the only stationary, axisymmetric, asymptotically-NHEK solutions with smooth horizons are diffeomorphic to NHEK.  Since we expect charges to be captured by highly symmetric solutions, this result strongly suggests that dynamics in the NHEK background are
subject to linearization-stability constraints associated with the full set of $SL(2,\R) \times U(1)$ isometries.  Considering simple scaling limits of perturbed asymptotically flat Kerr black holes lent additional support to this hypothesis.  Subtleties involving boundary gravitons were discussed in section \ref{LinStab}.

We then explored scalar and tensor perturbations in sections \ref{scalars} and \ref{grav}, finding that the linearization-stability constraints greatly restricted the solutions.  In particular, we saw that generalized Dirichlet boundary conditions for scalars or Teukolsky-Dirichlet boundary conditions for tensors were consistent only with  trivial solutions: $\Phi = 0$,  or $h=0$ up to linearized diffeomorphisms.  
It remains possible that some more general set of boundary conditions allowing the fields to fall-off more slowly at infinity would allow non-trivial solutions, though we consider this unlikely.

However, we did find a large family of solutions satisfying Teukolsky-Dirichlet boundary conditions which solved all constraints at $t=0$.  The problem was that boundary conditions conserving symplectic flux (Klein-Gordon flux for scalars) tended not to preserve the full $SL(2,\R)$ symmetry.  As a result, certain $SL(2,\R)$ charges were not naturally conserved.  Requiring the associated charges to vanish at all times was thus a much stronger constraint than just imposing them at $t=0$, leading to the paucity of solutions described above.  We noted that this amounts to a failure of the Cauchy problem for such boundary conditions at the non-linear level. 

It is interesting to reflect on the implications for the conjectured Kerr/CFT correspondence of \cite{Guica:2008mu}.    Before doing so, however, we must reconcile the boundary conditions used in various parts of this work with those used in \cite{Guica:2008mu}.   While our basic scaling arguments (section \ref{limits}) did not rely on any particular boundary conditions, the uniqueness theorem for stationary axisymmetric asymptotically NHEK solutions with smooth horizons required that the metric approach (\ref{nhekscaling}) at large $r$.  In contrast, the fall-off conditions of \cite{Guica:2008mu} (which we call GHSS fall-off) allow departures in the leading terms of certain components of the metric.  While such cases were not included in our analysis, we believe that a similar uniqueness theorem should nevertheless hold.  In particular, recall that appendix A of \cite{Guica:2008mu} studied precisely these departures from (\ref{nhekscaling}) in the limit where they are small.  The linearized Einstein equations then implied that such terms were determined by a single function $f(t, \phi)$.  In order for $\partial_t$ and $\partial_\phi$ to remain symmetries, this function must be constant.  But one may also show that any non-zero constant forces the energy to diverge.  While it remains to perform a complete non-linear analysis, we take this as evidence that GHSS fall-off allows no new stationary axisymmetric solutions.  While \cite{Guica:2008mu}  also imposed the constraint $Q_{\partial_t} =0$, we  expect that no generalization to the case $Q_{\partial_t} \neq 0$ is possible.  We also expect the other $SL(2,\mathbb{R})$ charges to vanish for all smooth solutions consistent with GHSS fall-off.

As a result, we are led to the same linearization-stability constraints studied in sections \ref{scalars} and \ref{grav}, but now subject to boundary conditions implied by GHSS fall-off.  In \ref{grav}, we argued that  the constraints admit no non-trivial solutions, up to linearized diffeomorphisms.  These arguments were definitive for boundary conditions that allow only modes with $\Re[\mu] \le  0$.  Now, it is clear from (\ref{Tpert}) that, at least as written in our in-going gauge, modes  $h_{ab}^{(\mu)}$ are only compatible with GHSS fall-off when $\Re[\mu]$ is sufficiently negative.  But one can nevertheles ask if modes with $\Re[\mu] > 0$ might be made compatible by the application of a linearized diffeomorphism.   It turns out that this is not possible, as can be shown by using the fact that $\psi_0, \psi_4$ are invariant under linearized diffeomorphisms.  Evaluating $\psi_0, \psi_4$ for GHSS fall-off gives behavior inconsistent with that of $h_{ab}^{(\mu)}$  for $\Re[\mu] >  0$ (in fact for $\Re[\mu] >  -1/2$).    Thus, only pure linearized diffeomorphisms can satisfy the constraints at all times.  

This makes the situation similar to that of chiral gravity \cite{Maloney:2009ck}, in that the boundary conditions remove potential instabilities.  On the other hand, there are also two significant differences from the chiral gravity case:  First, the price of removing these instabilities is an apparent lack of a good Cauchy problem (see again footnote \ref{extend} from section \ref{Tcharges}).       Second, in contrast to the fact that chiral gravity admits BTZ black holes, in the present context  all stationary axisymmetric black hole solutions are diffeomorphic to the original extreme throat.   As a result, it is indeed natural to consider the NHEK geometry to be a ground state as suggested in \cite{Guica:2008mu}.  It will be interesting to see whether these features play a role in future Kerr/CFT developments.

Finally, the reader may wonder how our results generalize to spacetimes constructed from near-horizon limits of other rotating black holes.  For definiteness, we confine our comments to the 3+1 Kerr-Newman case.  The near-horizon limit of extreme Kerr-Newman for general angular momentum $J$ and charge $Q$ was analyzed in \cite{Bardeen:1999px,Hartman:2008pb} where it was found that the asymptotic structure is very similar to that of (\ref{nhekmetric}).   We therefore expect a similar set of linearization-stability constraints.  One difference, however, is that when $J/Q^2$ becomes smaller than the critical value $2/3$ (i.e., close enough to the Reissner-Nordstrom solution), the velocity of light surface in the asymptotically flat extreme black hole detaches from the horizon.  As a result, for small $J/Q^2$ the near-horizon solution {\it does} have a globally timelike or null Killing field.  We therefore expect no negative energy perturbative excitations for uncharged scalars and gravitons, so that imposing even just the constraint $Q_{\partial_t}=0$ at a single time should forbid all linearized solutions.  This is not yet the end of the story, however, as Einstein-Maxwell theory has extreme black holes.  An effective quantum description of these black holes should involve charged fields (with charges $q = m$).  Since $q = m$ scalar fields exhibit superradiance near any extreme Kerr-Newman black hole with $J \neq 0$, we would expect the inclusion of quantum effects involving extreme black holes to make all cases with $J \neq 0$ similar to that of the $Q=0$ extreme Kerr throat analyzed in this work.

\medskip
{\it Note Added:} During the completion of this work, we learned of \cite{Harvey}, which has some overlap with our discussion above.

\section*{Acknowledgements}
It is a pleasure to thank Geoffrey Comp\`ere for many useful discussions during the course of the project.  This work was supported in part by the US National Science Foundation under Grant No.~PHY05-55669, and by funds from the University of California.

\appendix
\section{The Newman-Penrose tetrad and the NHEK geometry}
\label{NP}
A Newman-Penrose tetrad \cite{Newman:1961qr} consists of two real null vectors $l,~n$, and one complex null vector $m$ satisfying\footnote{Our definitions for the Newman-Penrose formalism are consistent with $(-+++)$ metric signature (see e.g., \cite{Frolov:1998wf}).  Note that there are thus certain sign differences with respect to the definitions in \cite{Teukolsky:1973ha}, as Teukolsky works in  $(+---)$ signature.}
\be
l\cdot n=-m\cdot m^* =-1 \,.
\ee
All other inner products are zero. It follows that the inverse metric can be expressed in the form
\be
g^{ab}=- l^an^b - n^al^b + m^am^{*b} + m^{*a}m^b \,.
\ee
It is convenient to define a set of differential operators given by taking partial derivatives in the tetrad directions:
\be
D=l^a\frac{\partial}{\partial x^a},~\Delta=n^a\frac{\partial}{\partial x^a},~\delta=m^a\frac{\partial}{\partial x^a}~,\delta^*=m^{*a}\frac{\partial}{\partial x^a}.
\ee
The connection is expressed in terms of the ``spin coefficients," which are defined as
\bea
\nonumber-\kappa=l_{a;b}m^al^{b} ,&\nu=n_{a;b}m^{*a}m^b,\\
\nonumber-\rho=l_{a;b}m^am^{*b},&\mu=n_{a;b}m^{*a}{m^b},\\
\nonumber-\sigma=l_{a;b}m^am^b,&\lambda=n_{a;b}m^{*a}m^{*b},\\
\nonumber-\tau=l_{a;b}m^an^b,&\pi=n_{a;b}m^{*a}l^b,\\
\nonumber-\epsilon=\frac{1}{2}(l_{a;b}n^al^b-m_{a;b}m^{*a}l^b),&-\gamma=\frac{1}{2}(l_{a;b}n^an^b-m_{a;b}m^{*a}n^b),\\
-\alpha=\frac{1}{2}(l_{a;b}n^am^{*b}-m_{a;b}m^{*a}m^{*b}),&-\beta=\frac{1}{2}(l_{a;b}n^am^b-m_{a;b}m^{*a}m^b).
\eea
Here the semi-colon denotes a covariant derivative, e.g. $l_{a;b} = \nabla_b l_a$.
The Newman-Penrose Weyl scalars are given by certain components of the Weyl tensor:
\bea
\nonumber\psi_0=C_{abcd}l^am^bl^cm^d,&&~\psi_1=C_{abcd}l^an^bl^cm^d,\\
\nonumber&&\hspace{-3.5cm}\psi_2=\frac{1}{2}C_{abcd}(l^an^bl^cn^d-l^an^bm^cm^{*d}),
\\
\psi_3=-C_{abcd}l^an^bn^cm^{*d},&&~\psi_4=C_{abcd}n^am^{*b}n^cm^{*d}.
\eea
For Kerr spacetime in Boyer-Lindquist coordinates $(\th,\rh,\theta,\phih)$, the ``Kinnersley tetrad'' is
\bea
\nonumber l^a&=&[(\rh^2+a^2)/\Delta,1,0,a/\Delta],\\
n^a&=&[\rh^2+a^2,-\Delta,0,a]/(2\Sigma),\\
m^a&=&[i a \sin\theta,0,1,i \csc\theta]/\sqrt{2}(\rh+i a \cos\theta),
\eea
where $\Delta=\rh^2-2M\rh+a^2$ and $\Sigma=\rh^2+a^2\cos^2\theta$.

Under the coordinate change (\ref{nhekscaling}) we find $l\propto 1/\lambda$, $n\propto \lambda$, and so we require a tetrad rotation $l\rightarrow \lambda l,~n\rightarrow n/\lambda$ before taking the limit $\lambda \to 0$. Thus, in the coordinates $(t,r,\theta,\phi)$ (with $2 M^2 = 1$), we take the null tetrad in the NHEK geometry to be
\bea
\nonumber l^a&=&[1/r^2,1,0,-1/r],\\
\nonumber n^a&=&[1/(1+\cos^2\theta),-r^2/(1+\cos^2\theta),0,-r/(1+\cos^2\theta)],\\
m^a&=&[0,0,-i/(\cos\theta-i),(\cos\theta+i)/(2\sin\theta)].
\eea
The non-zero spin coefficients for the NHEK spacetime are then
\begin{eqnarray}
\nonumber
\beta =  \frac{\cot\theta}{2(1+i \cos \theta)},&& \pi = \frac{i \sin \theta}{1- i \cos \theta}, \qquad \tau =- \frac{i \sin \theta}{1+ \cos^2 \theta},\\
&&\hspace{-1.3cm}\gamma = \frac{r}{1+\cos^2 \theta},\quad \alpha = \pi-\beta^* \,.
\end{eqnarray}
The Weyl scalars are $\psi_0 = \psi_1 = \psi_3 = \psi_4 = 0$ and
\begin{equation}
\psi_2=-\frac{2}{(1- i \cos \theta)^3} \,.
\end{equation}

\setcounter{equation}{0}
\section{Scalars in global coordinates}
\label{global}
This appendix briefly summarizes  the behavior of massless scalars in the global NHEK geometry. Since there are two boundaries (at $y=\pm\infty$), we must impose two boundary conditions.  This will lead to a quantized spectrum.

The wave equation again separates in global coordinates.  One finds solutions $\Phi=e^{-i \omega \tau}e^{+im\varphi}\Theta(\theta)Y(y)$ where $\Theta(\theta)$ satisfies (\ref{NHEKscalarangular}) and where $Y(y)$ satisfies
\be
(fY')'+\left(\frac{1}{4}-\mu^2-m^2+\frac{(\omega+m y)^2}{1+y^2} \right)Y=0.  \label{globalY}
\ee
As usual,  $2m^2-K=1/4-\mu^2$.  The asymptotics of (\ref{globalY}) agree with (\ref{NHEKwave}) and we again have $Y\sim y^{-1/2\pm\mu}$ .

Eqn. (\ref{globalY}) has regular singular points at $y=\pm i,\infty$ and has solutions in terms of hypergeometric functions. Written in terms of $z=\frac{1+iy}{2}$, we have
\bea
Y=C_1 z^\frac{im+\omega}{2}(z-1)^\frac{im-\omega}{2}{}_2F_1(\frac{1}{2}+im-\mu,\frac{1}{2}+im+\mu,1+im+ \omega;z)&\nonumber\\
+C_2z^\frac{-im-\omega}{2}(z-1)^\frac{im-\omega}{2}{}_2F_1(\frac{1}{2}-\mu-\omega,\frac{1}{2}+\mu-\omega,1-im-\omega;z) .&
\eea
Consider a solution of the form $A y^{-1/2+\mu}+B y^{-1/2-\mu}$ near $y=+\infty$. It is a straightforward but tedious calculation to verify that near $y=-\infty$ the solution is of the form $\tilde{A} y^{-1/2+\mu}+\tilde{B} y^{-1/2-\mu}$, where

\begin{figure}
\begin{center}
\includegraphics[width=5.75cm]{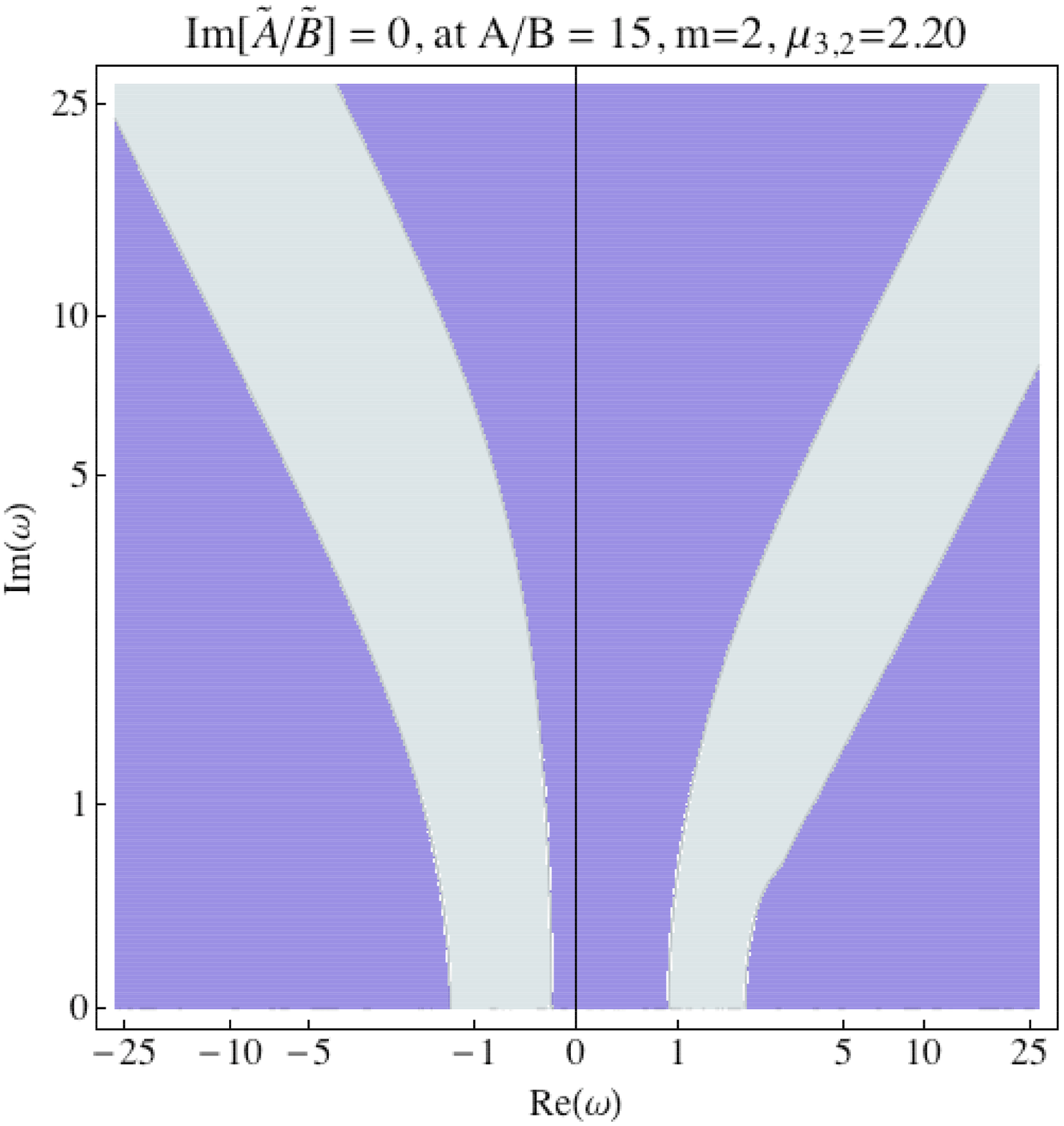}
\includegraphics[width=5.75cm]{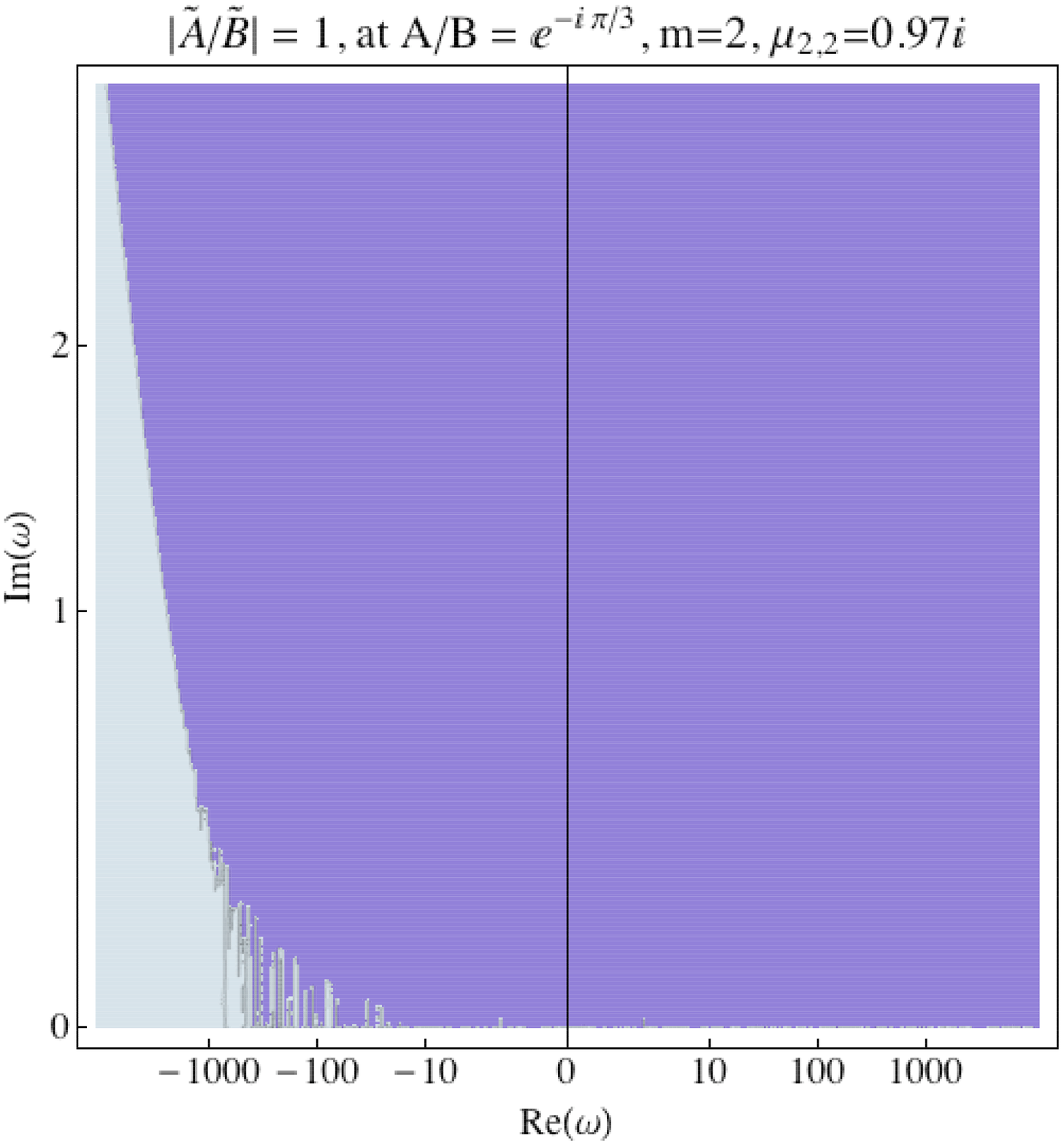}
\end{center}
\begin{caption}
{Typical plots of $\Im[\tilde A/\tilde B]$ for power law modes (left) and $|\tilde A/\tilde B|  - 1$ for oscillatory modes (right) over the complex $\omega$ plane. The shading denotes positive/negative values.  Frequencies that satisfy boundary conditions which conserve Klein-Gordon flux at the $y=+\infty$ boundary lie on the boundaries.  For each plot, $A/B$ has already been fixed to conserve flux at the other boundary ($y=-\infty$). \label{Global}}
\end{caption}
\end{figure}

\be
\frac{\tilde{A}}{\tilde{B}}=\frac{a_1 A + a_2 B}{b_1 B + b_2 A},\ee
for

\bea
a_1&=&-\frac{\pi^3e^{-i\pi\mu}\csc(2\pi\mu)}{2}(e^{2m\pi}\sec\pi(\mu+\omega)\sech\pi(m-i\mu)\nonumber\\
&&-e^{2i\pi\omega}\sec\pi(\mu-\omega)\sech\pi(m+i\mu) ),\nonumber\\
a_2&=&-\mu4^{-\mu}(e^{2m\pi}-e^{2\pi i\omega})\Gamma(2\mu)^2\Gamma(\frac{1}{2}-i m-\mu) \Gamma (\frac{1}{2}+i m-\mu)\nonumber\\
&&\times\Gamma(\frac{1}{2}-\mu -\omega) \Gamma(\frac{1}{2}-\mu +\omega)\nonumber\\
b_1&=&a_1(\mu\rightarrow -\mu),\nonumber\\
b_2&=&a_2(\mu\rightarrow -\mu).
\eea

Since the asymptotics are the same as in the Poincar\'e case, imposing conservation of Klein-Gordon flux at each boundary leads to familiar results.  Boundary conditions which leave modes with different $(\ell,m)$ uncoupled take the form:
\begin{eqnarray}
{\rm power \  law \  modes:} & \    A_{\ell, m} /B_{\ell, m} =\alpha_{\ell, m} ,~\tilde{A}_{\ell, m} /\tilde{B}_{\ell, m} =\tilde{\alpha}_{\ell, m} ,~\alpha_{\ell, m} ,\tilde{\alpha}_{\ell, m} \in \mathbb{R} \cr
{\rm oscillatory \  modes:} &  \
A_{\ell, m} /B_{\ell, m} =e^{i\beta_{\ell, m} },~\tilde{A}_{\ell, m} /\tilde{B}_{\ell, m} =e^{i\tilde{\beta}_{\ell, m} },~\beta_{\ell, m} ,\tilde{\beta}_{\ell, m} \in\mathbb{R}.
\end{eqnarray}

These conditions are difficult to analyze analytically, but fixing $\alpha, \beta$, one can numerically solve for the curves in the complex $\omega$ plane where $\Im(\tilde{A}/\tilde{B})=0$ for power-law modes and $|\tilde{A}/\tilde{B}|=1$ for oscillatory modes.  Some typical results are shown in figure \ref{Global}.  Choosing particular values of $\tilde \alpha, \tilde  \beta$ then selects a discrete set of frequencies along this curve.


\begin{thebibliography}{99}

\bibitem{Zaslavskii:1997uu}
  O.~B.~Zaslavskii,
  ``Horizon/Matter Systems Near the Extreme State,''
  Class.\ Quant.\ Grav.\  {\bf 15} (1998) 3251
  [arXiv:gr-qc/9712007].


\bibitem{Bardeen:1999px}
 J.~M.~Bardeen and G.~T.~Horowitz,
 ``The extreme Kerr throat geometry: A vacuum analog of AdS(2) x S(2),''
 Phys.\ Rev.\  D {\bf 60} (1999) 104030
 [arXiv:hep-th/9905099].


\bibitem{Guica:2008mu}
 M.~Guica, T.~Hartman, W.~Song and A.~Strominger,
 ``The Kerr/CFT Correspondence,''
 arXiv:0809.4266 [hep-th].

\bibitem{Maldacena:1997re}
 J.~M.~Maldacena,
 ``The large N limit of superconformal field theories and supergravity,''
 Adv.\ Theor.\ Math.\ Phys.\  {\bf 2} (1998) 231
 [Int.\ J.\ Theor.\ Phys.\  {\bf 38} (1999) 1113]
 [arXiv:hep-th/9711200].

\bibitem{Azeyanagi:2009wf}
T.~Azeyanagi, G.~Comp\`ere, N.~Ogawa, Y.~Tachikawa and S.~Terashima,
  ``Higher-Derivative Corrections to the Asymptotic Virasoro Symmetry of 4d
  Extremal Black Holes,''
  arXiv:0903.4176 [hep-th].


\bibitem{SuperRad}
Ya.B. Zel'dovich,  ``Amplification of Cylindrical Electromagnetic Waves Reflected from a Rotating Body," JETP Lett. 14 (1971) 180;
C.W. Misner, Bull. Am. Phys. Soc. 17 (1972) 472;
J.M. Bardeen, S.A. Teukolsky, and W.H. Press, ``Rotating Black Holes: Locally Nonrotating Frames, Energy Extraction, and Scalar Synchrotron Radiation," Astrophys. J., 178, 347 (1972);
A. A. Starobinsky, ``Amplification of Waves during Reflection from a Rotating Black Hole," Zh. Eksp. Teor. Fiz 64, 48 (1973) [Sov.
Phys. JETP 37, 28 (1973)].

\bibitem{PTBomb} W.H. Press and S.A. Teukolsky,
``Floating Orbits, Superradiant Scattering and the Black-Hole Bomb,'' Nature, 238 (1972) 211.


\bibitem{Cardoso:2004nk}
 V.~Cardoso, O.~J.~C.~Dias, J.~P.~S.~Lemos and S.~Yoshida,
 ``The black hole bomb and superradiant instabilities,''
 Phys.\ Rev.\  D {\bf 70} (2004) 044039
 [Erratum-ibid.\  D {\bf 70} (2004) 049903]
 [arXiv:hep-th/0404096].

\bibitem{Frolov:1989jh}
 V.~P.~Frolov and K.~S.~Thorne,
 ``Renormalized Stress-Energy Tensor near the Horizon of a Slowly Evolving, Rotating Black Hole,''
 Phys.\ Rev.\  D {\bf 39} (1989) 2125.


\bibitem{Kay:1988mu}
B.~S.~Kay and R.~M.~Wald,
``Theorems on the Uniqueness and Thermal Properties of Stationary, Nonsingular, Quasifree States on Space-Times with a Bifurcate Killing Horizon,''
 Phys.\ Rept.\  {\bf 207} (1991) 49.


\bibitem{Ottewill:2000qh}
 A.~C.~Ottewill and E.~Winstanley,
 ``The renormalized stress tensor in Kerr space-time: General results,''
 Phys.\ Rev.\  D {\bf 62} (2000) 084018
 [arXiv:gr-qc/0004022].

\bibitem{Ottewill:2000yr}
 A.~C.~Ottewill and E.~Winstanley,
 ``Divergence of a quantum thermal state on Kerr space-time,''
 Phys.\ Lett.\  A {\bf 273} (2000) 149
 [arXiv:gr-qc/0005108].




\bibitem{BD} D. Brill and S. Deser, ``Instability of Closed Spaces in General Relativity," Commun. Math. Phys. {\bf 32} (1973) 291.

\bibitem{FM1} A. Fischer and J. Marsden, ``Linearization Stability of the Einstein Equations," Bull. Amer. Math. Soc. {\bf 79} (1973) 997.


\bibitem{VMLS1} V. Moncrief, ``Space-time symmetries and linearization stability of the Einstein equation, I,"  J. Math Phys. {\bf 16} (1975) 493.

\bibitem{VMLS2} V. Moncrief, ``Space-time symmetries and linearization stability of the Einstein equations. II,'' J. Math Phys. {\bf 17} (1976) 1893.

\bibitem{Arms} J. Arms, ``Linearization stability of the Einstein-Maxwell system,'' J. Math. Phys 18 (1977) 830; ``Linearization stability of gravitational and gauge fields,'' J. Math. Phys. 20 (1979) 443.

\bibitem{FM} A. E. Fisher and J. E. Marsden, in {\it General Relativity: An Einstein Centenary Survey}, edited by S. W. Hawking and W. Israel (Cambridge University Press, Cambridge, England, 1979).

\bibitem{AMM}  J. Arms, J. Marsden and V. Moncrief, ``The Structure of the Space of Solutions of Einstein's Equations II,''  Annals of Physics 144 (1982) 81.


\bibitem{Maloney:2009ck}
 A.~Maloney, W.~Song and A.~Strominger,
 ``Chiral Gravity, Log Gravity and Extremal CFT,''
 arXiv:0903.4573 [hep-th].
 
\bibitem{Teukolsky:1973ha}
 S.~A.~Teukolsky,
 ``Perturbations of a rotating black hole. 1. Fundamental equations for
 gravitational electromagnetic and neutrino field perturbations,''
 Astrophys.\ J.\  {\bf 185} (1973) 635.


\bibitem{Press:1973zz}
W.~H.~Press and S.~A.~Teukolsky,
 ``Perturbations of a Rotating Black Hole. II. Dynamical Stability of the Kerr
 Metric,''
 Astrophys.\ J.\  {\bf 185} (1973) 649.

\bibitem{Teukolsky:1974yv}
 S.~A.~Teukolsky and W.~H.~Press,
 ``Perturbations Of A Rotating Black Hole. III - Interaction Of The Hole With
 Gravitational And Electromagnet Ic Radiation,''
 Astrophys.\ J.\  {\bf 193} (1974) 443.



\bibitem{Newman:1961qr}
 E.~Newman and R.~Penrose,
 ``An Approach to gravitational radiation by a method of spin coefficients,''
 J.\ Math.\ Phys.\  {\bf 3} (1962) 566.




\bibitem{Maldacena:1998uz}
 J.~M.~Maldacena, J.~Michelson, and A.~Strominger,
 ``Anti-de Sitter fragmentation,''
 JHEP {\bf 9902} (1999) 011
 [arXiv:hep-th/9812073].



\bibitem{Barnich:2001jy}
 G.~Barnich and F.~Brandt,
 ``Covariant theory of asymptotic symmetries, conservation laws and  central
 charges,''
 Nucl.\ Phys.\  B {\bf 633} (2002) 3
 [arXiv:hep-th/0111246].


\bibitem{Hollands:2007aj}
  S.~Hollands and S.~Yazadjiev,
  ``Uniqueness theorem for 5-dimensional black holes with two axial Killing
 fields,''
  Commun.\ Math.\ Phys.\  {\bf 283} (2008) 749
  [arXiv:0707.2775 [gr-qc]].

\bibitem{Hollands:2008fm}
  S.~Hollands and S.~Yazadjiev,
 ``A uniqueness theorem for stationary Kaluza-Klein black holes,''
  arXiv:0812.3036 [gr-qc].

\bibitem{Mazur:1982db}
  P.~O.~Mazur,
  ``Proof Of Uniqueness Of The Kerr-Newman Black Hole Solution,''
  J.\ Phys.\ A  {\bf 15} (1982) 3173.

\bibitem{Chrusciel:2008js}
  P.~T.~Chrusciel and J.~Lopes Costa,
  ``On uniqueness of stationary vacuum black holes,''
  arXiv:0806.0016 [gr-qc].

  \bibitem{us}  A.~J.~Amsel, G.~T.~Horowitz, D.~Marolf, and M.~M.~Roberts, ``Uniqueness of Extremal Kerr and Kerr-Newman Black Holes,'' arXiv:0906.2367 [gr-qc].

  \bibitem{Weinstein}
  See Lemma 8 in: G. Weinstein, ``On the Dirichlet problem for harmonic maps with prescribed singularities", Duke Math. J. {\bf 77} (1995) 135.

\bibitem{Kunduri:2008rs}
  H.~K.~Kunduri and J.~Lucietti,
  ``A classification of near-horizon geometries of extremal vacuum black
  holes,''
  arXiv:0806.2051 [hep-th].
  H.~K.~Kunduri and J.~Lucietti,
``Uniqueness of near-horizon geometries of rotating extremal AdS(4) black
  holes,''
  Class.\ Quant.\ Grav.\  {\bf 26} (2009) 055019
  [arXiv:0812.1576 [hep-th]].


\bibitem{Corvino:2003sp}
J.~Corvino and R.~M.~Schoen,
 ``On the Asymptotics for the Vacuum Einstein Constraint Equations,''
 arXiv:gr-qc/0301071.

\bibitem{MTW}
C. W. Misner, K. S. Thorne, and J. A. Wheeler, {\it Gravitation}, (New York, W H. Freeman \& Co., 1973).




\bibitem{Breitenlohner:1982bm}
  P.~Breitenlohner and D.~Z.~Freedman,
  ``Positive Energy In Anti-De Sitter Backgrounds And Gauged Extended
  Supergravity,''
  Phys.\ Lett.\  B {\bf 115} (1982) 197;
  P.~Breitenlohner and D.~Z.~Freedman,
  ``Stability In Gauged Extended Supergravity,''
  Annals Phys.\  {\bf 144} (1982) 249.

  \bibitem{gr}
I.~S.~Gradshteyn and I.~M.~Ryzhik,
{\it Table of Integrals, Series, and Products}, San Diego, Academic Press, Inc., (1980).

\bibitem{Ishibashi:2004wx}
A.~Ishibashi and R.~M.~Wald,
  ``Dynamics in non-globally hyperbolic static spacetimes. III: anti-de  Sitter
  spacetime,''
  Class.\ Quant.\ Grav.\  {\bf 21} (2004) 2981
  [arXiv:hep-th/0402184].

\bibitem{nonrotating}
  T.~Regge and J.~A.~Wheeler,
  ``Stability Of A Schwarzschild Singularity,''
  Phys.\ Rev.\  {\bf 108} (1957) 1063;
  C.~V.~Vishveshwara,
  ``Stability of the schwarzschild metric,''
  Phys.\ Rev.\  D {\bf 1} (1970) 2870;
  F.~J.~Zerilli,
  ``Gravitational field of a particle falling in a schwarzschild geometry
  analyzed in tensor harmonics,''
  Phys.\ Rev.\  D {\bf 2} (1970) 2141;
  ``Effective Potential For Even Parity Regge-Wheeler Gravitational
  Perturbation Equations,''
  Phys.\ Rev.\ Lett.\  {\bf 24} (1970) 737.


\bibitem{wald:1453}
Robert~M.~Wald,
``On perturbations of a Kerr black hole,''
J. Math. Phys. {\bf 14} 10 (1973) 1453-1461.

\bibitem{Detweiler:1980gk}
  S.~Detweiler,
  ``Black Holes And Gravitational Waves. Iii. The Resonant Frequencies Of
  Rotating Holes,''
  Astrophys.\ J.\  {\bf 239} (1980) 292.

\bibitem{Hod:2008zz}
  S.~Hod,
  ``Slow relaxation of rapidly rotating black holes,''
  Phys.\ Rev.\  D {\bf 78} (2008) 084035
  [arXiv:0811.3806 [gr-qc]].

\bibitem{Chrzanowski:1975wv}
 P.~L.~Chrzanowski,
 ``Vector Potential And Metric Perturbations Of A Rotating Black Hole,''
 Phys.\ Rev.\  D {\bf 11} (1975) 2042.





\bibitem{Wald:1978vm}
 R.~M.~Wald,
 ``Construction Of Solutions Of Gravitational, Electromagnetic, Or Other
 Perturbation Equations From Solutions Of Decoupled Equations,''
 Phys.\ Rev.\ Lett.\  {\bf 41} (1978) 203.

\bibitem{Barnich:2007bf}
 G.~Barnich and G.~Comp\`ere,
 ``Surface charge algebra in gauge theories and thermodynamic integrability,''
 J.\ Math.\ Phys.\  {\bf 49} (2008) 042901
 [arXiv:0708.2378 [gr-qc]].

\bibitem{Wald:1999wa}
 R.~M.~Wald and A.~Zoupas,
 ``A General Definition of ``Conserved Quantities'' in General Relativity and
 Other Theories of Gravity,''
 Phys.\ Rev.\  D {\bf 61} (2000) 084027
 [arXiv:gr-qc/9911095].

\bibitem{Compere:2008us}
 G.~Comp\`ere and D.~Marolf,
 ``Setting the boundary free in AdS/CFT,''
 Class.\ Quant.\ Grav.\  {\bf 25} (2008) 195014
 [arXiv:0805.1902 [hep-th]].

\bibitem{Hartman:2008pb}
  T.~Hartman, K.~Murata, T.~Nishioka and A.~Strominger,
  ``CFT Duals for Extreme Black Holes,''
  JHEP {\bf 0904}, (2009) 019
  [arXiv:0811.4393 [hep-th]].


\bibitem{Harvey} Oscar~J.C.~Dias, Harvey~S. Reall, Jorge~E.~Santos,
``Kerr-CFT and gravitational perturbations'',
arXiv:0906.2380 [hep-th].

\bibitem{Frolov:1998wf}
  V.~P.~Frolov and I.~D.~Novikov,
  {\it Black hole physics: Basic concepts and new developments},
Dordrecht, Netherlands: Kluwer Academic (1998).



\end{thebibliography}
\end{document}